\documentclass[twocolumn]{aastex63}

\newcommand{\sextractor}{\texttt{SExtractor}}
\newcommand{\scamp}{\texttt{SCAMP}}
\newcommand{\swarp}{\texttt{SWarp}}

\newcommand{\psfex}{\texttt{PSFEx}}

\newcommand{\g}{$g$}%
\newcommand{\ib}{$i$}%
\definecolor{purple}{rgb}{0.5,0,0.87}
\definecolor{teal}{rgb}{0,0.75,0.75}
\hypersetup{linkcolor=teal,citecolor=purple,filecolor=cyan,urlcolor=teal}

\received{}
\revised{}
\accepted{}
\submitjournal{ApJ}

\shorttitle{The buildup of the ICL of A85 with HSC}
\shortauthors{Montes et al.}
\graphicspath{{./}{figures/}}

\begin{document}

\title{The buildup of the intracluster light of Abell 85 as seen by Subaru's Hyper Suprime-Cam}

\correspondingauthor{Mireia Montes}
\email{mireia.montes.quiles@gmail.com}

\author[0000-0001-7847-0393]{Mireia Montes}
\affiliation{School of Physics, University of New South Wales, Sydney, NSW 2052, Australia}
\author[0000-0002-9796-1363]{Sarah Brough}
\affiliation{School of Physics, University of New South Wales, Sydney, NSW 2052, Australia}
\author[0000-0002-2879-1663]{Matt S. Owers}
\affiliation{Department of Physics and Astronomy, Macquarie University, NSW 2109, Australia}
\affiliation{Astronomy, Astrophysics and Astrophotonics Research Centre, Macquarie University, Sydney, NSW 2109, Australia}
\author[0000-0003-3283-4686]{Giulia Santucci}
\affiliation{School of Physics, University of New South Wales, Sydney, NSW 2052, Australia}



\begin{abstract}
The study of low surface brightness light in large, deep imaging surveys is still uncharted territory as automated data reduction pipelines over-subtract or eliminate this light. Using archival data of the Abell 85 cluster of galaxies taken with Hyper Suprime-Cam on the Subaru Telescope, we show that using careful data processing can unveil the diffuse light within the cluster, the intracluster light. We reach surface brightness limits of $\mu_{g}^{limit}$(3$\sigma$, $10\arcsec\times10\arcsec$) = 30.9 mag/arcsec$^2$, and $\mu_{i}^{limit}$(3$\sigma$, $10\arcsec\times10\arcsec$) = 29.7 mag/arcsec$^2$. 
We measured the radial surface brightness profiles of the brightest cluster galaxy out to the intracluster light (radius $\sim215$ kpc), for the $g$ and $i$ bands. We found that both the surface brightness and the color profiles become shallower beyond $\sim75$ kpc suggesting that a distinct component, the intracluster light, starts to dominate at that radius. The color of the profile at $\sim100$ kpc suggests that the buildup of the intracluster light of Abell 85 occurs by the stripping of massive ($\sim10^{10}M_{\odot}$) satellites. The measured fraction of this light ranges from $8\%$ to $30\%$ in \g{}, depending on the definition of intracluster light chosen.
\end{abstract} 




\section{Introduction}
Deep observations of galaxy clusters have revealed the existence of a diffuse glow produced by stars not bound to any individual galaxy; the intracluster light \citep[ICL, see][for reviews]{Mihos2016, Montes2019}. As the by-product of galaxy interactions, the ICL forms a fossil record of all the dynamical processes the system has undergone and provides a holistic view of the interaction history of the cluster \citep[e.g.,][]{Merritt1984}. 

The ICL is key to understanding how brightest cluster galaxies (BCGs) grow with time. Their formation and evolution have been predicted to be rather different than satellite galaxies \citep[e.g.,][]{DeLucia2007}. The innermost regions of these massive galaxies appear to have formed the majority of their stars at high redshift and on short timescales \citep[e.g.,][]{Thomas2005} whereas their outer parts are likely assembled as a consequence of multiple minor merging more recently \citep[e.g.,][]{Trujillo2011}. As the ICL is often found to be more concentrated around the BCG \citep[e.g.,][]{Mihos2005}, it implies that the growth of both components, BCG and ICL, are connected. In addition, simulations of the growth rate of BCGs better agree with observations if formation of ICL is included \citep[e.g.,][]{Conroy2007, Contini2019, Spavone2020}.

A useful tool to characterize the ICL is the study of its stellar populations, as they reflect the properties of the galaxies from which the ICL accreted its stars. Knowing the stellar populations of the ICL in clusters allow us to infer the mechanisms at play in the formation of this component, and therefore how (and when) the assembly history of these clusters occurred. Observations show clear radial gradients in colors indicating radial gradients in metallicity \citep[e.g.,][]{Zibetti2005, Iodice2017, Mihos2017, DeMaio2015, DeMaio2018} and, in some cases, age \citep[e.g.,][]{MT14, Morishita2017, MT18}. These studies point to the tidal stripping of massive satellites \citep[a few $\times10^{10}M_{\odot}$;][]{MT14, MT18} as the dominant process of ICL formation for massive clusters ($\sim10^{15}M_\odot$)\footnote{Diffuse light has also been detected and studied in groups of galaxies \citep[e.g.,][]{DaRocha2005, DeMaio2018, Iodice2020} but the main mechanism of the formation of this intragroup light appears to differ from that for clusters \citep[e.g.,][]{Spavone2020}}. 

A clear limitation in the study of the ICL is the lack of statistically significant samples with the required depth \citep[$\mu_V > 26.5$ mag/arcsec$^2$;][]{Rudick2006}. Unfortunately, the long exposure times required for these observations mean that very few clusters have been studied, so far. To date, studies have only analysed small samples \citep[1-20 clusters; ][]{Krick2007, MT14, MT18, Burke2015, Jimenez-Teja2018} or employed stacking of many clusters to obtain a coarse measurement \citep[e.g.][]{Zibetti2005, Zhang2019}.

This is changing with the next generation of surveys using state-of-the-art cameras that will be able to reach unprecedented depths over large areas in the sky. An example is the Hyper Suprime-Cam \citep[HSC;][]{Miyazaki2018} on the 8.2-meter Subaru Telescope. This camera is well suited to not only provide the wide field-of-view necessary to observe nearby clusters but also the time efficiency of a large telescope being able to reach ICL depths in short exposure times. The HSC is currently carrying out the HSC Subaru Strategic Program (HSC-SSP), a survey of 1400 deg$^2$ in five different bands ($grizy$) plus four narrow filters. The depth and area of this survey will provide the large numbers of galaxy clusters necessary to \emph{deepen} our knowledge of the formation of the ICL \citep{Aihara2019}.

However, ICL studies need very accurate data processing. The data reduction of HSC data is undertaken with the HSC pipeline \citep{Bosch2018}, a custom version of the LSST\footnote{The Vera C. Rubin Observatory Legacy Survey for Space and Time.} pipeline. The sky subtraction algorithm in the HSC-SSP data release 1 over-subtracts extended halos of bright objects making it almost impossible to study nearby or very extended objects \citep{Aihara2018}\footnote{This issue was improved in the data release 2 \citep{Aihara2019} but not completely resolved.}. In addition, ICL studies are susceptible to biases due to flat-field inaccuracies and the scattered light from bright stars. 

In this work, we use archival HSC images of the cluster Abell 85 (A85) to test a dedicated data processing technique for low surface brightness science and study the diffuse light of this cluster out to $\sim 215$ kpc. The main properties of A85 are listed in Table \ref{tab:1}. A85 is a rich cluster of galaxies \citep[$\sim800$ spectrocopically confirmed galaxies within $2R_{200}$,][]{Owers2017, Habas2018} hosting a massive BCG (M$_*\sim3\times10^{12}M_{\odot}$, \citealt{Mehrgan2019}).
Many studies have shown that this cluster is slowly accreting material through several ongoing mergers with, at least, two subclusters or groups of galaxies \citep{BravoAlfaro2009, Ichinohe2015, Owers2017}. In addition, models of the X-ray temperature across the cluster support the picture that A85 has undergone several small mergers in the past few billion years \citep{Durret2005, Ichinohe2015}.

This cluster provides an ideal target for a pilot study of the ICL using HSC and dedicated data processing techniques for low surface brightness science. Studying the properties of the ICL in this cluster will inform us of the ongoing processes shaping this cluster, and its BCG.

Throughout this work, we adopt a standard cosmological model with the following parameters: $H_0=70$ km s$^{-1}$ Mpc$^{-1}$, $\Omega_m=0.3$ and $\Omega_\Lambda=0.7$. All magnitudes are in the AB magnitude system.

 \begin{deluxetable*}{lccccccc}[t]
\tablecaption{\label{tab:1}
Main properties of A85. Redshift, mass and radius are taken from \citet{Owers2017}.}
\tablehead{%
Name	        	& RA	      & DEC	     & z   & Distance  & Angular scale   & Virial M$_{200}$  & R$_{200}$     \\
                 &  [deg]     & [deg]    &    & [Mpc]  & [kpc/arcsec] & [10$^{14}M_\odot$] & [Mpc] }
\startdata
Abell 85         & 10.458750 & -9.301944 & 0.0549 & 245   & 1.068		&17.0$\pm$1.3 & $2.42$ 
\enddata
\end{deluxetable*}

\section{Data}

HSC is a 1.77 $\deg^2$ imaging camera on the Subaru Telescope operated by the National Astronomical Observatory of Japan (NAOJ) on the summit of Maunakea in Hawaii. It consists of 116 CCDs (104 science CCDs, 4 guide and 8 focus sensors) with a resolution of $0\farcs168$/pixel. For this work, we have used archival data. A85 was observed on the 2014-09-24 (Proposal ID: o14171). The science data consist of 9 frames in both the HSC-$G$ (\g) and the HSC-$I$ (\ib) bands. The exposure times for each frame are 200s and 240s, respectively. The observational strategy consisted of a dithering pattern of 9 positions around the center of the cluster. The offsets of $1\farcm367$ are enough to fill the gaps of the camera mosaic.
All the data used in this work was downloaded from the Subaru-Mitaka-Okayama-Kiso Archive \citep[SMOKA; ][]{Baba2002}\footnote{\url{https://smoka.nao.ac.jp/}}.

\subsection{Custom-made processing}
Exploring the ICL of clusters of galaxies is difficult as it is not only faint, but also extended. This means that in order to avoid biases when measuring the ICL caused by inhomogeneities in the images such as gradients and oversubtraction, see \citealt{Mihos2019} for a detailed description of the possible biases in low surface brightness imaging), the images must have a flat background and the background subtraction should be performed carefully so as not to eliminate this light.
At the time we started this project, the data reduced with the HSC pipeline \citep{Bosch2018} for the DR1 of the HSC-SSP survey \citep{Aihara2018}, showed significant oversubtraction around bright sources caused by a background estimation using a relatively small mesh size ($128\times128$ pix$^2 = \, 21\arcsec\times21\arcsec$). Because the cluster of interest is at low redshift (i.e, extended in the sky, $R_{200} = 2.42$ Mpc $= 0\fdg63 $; \citealt{Owers2017}), this oversubtraction would likely eliminate the ICL. 
For this reason, we developed a custom-made process in order to reduce the data, preserving low surface brightness light, i.e. the extended and faint ICL. The code is mainly written in Python and uses Astropy \citep{Astropy2018} and astronomical software such as \sextractor, \swarp, and \scamp \, \citep{Bertin1996, Bertin2002, Bertin2006}. The steps followed here to reduce the HSC images, after the individual CCD processing, are similar to those performed in \citet{Trujillo2016}.
For this work, as the images are dithered around the BCG of the cluster, we focus only on the innermost $40$ CCDs of the camera to reduce inaccuracies due to the non-uniform illumination of the CCDs. This corresponds to a radius of $\sim0 \fdg42$ ($1.6$ Mpc) around the BCG.
These are the main steps we conduct to process the data:

\begin{enumerate}
    \item Derivation of the calibration files (bias and dark)
    \item Individual CCD processing and assembly
    \item Flat-field derivation using science images and correction
    \item Camera mosaic with a careful determination of the sky background 
    \item Mosaic coaddition and final image calibration.
\end{enumerate}
In the following sections, we describe in detail how these steps are performed.


The HSC CCDs are full-depletion Hamamatsu CCDs \citep{Miyazaki2012}. The individual raw images are $2144\times4241$, divided into four science channels of $512\times4096$ pixels along with pre-scan, overscan and non-science regions of each of those channel.\footnote{As described in: \url{https://hsc.mtk.nao.ac.jp/pipedoc/pipedoc_4_e/e_hsc/index.html\#e-hsc}} Therefore, the next steps to calibrate each CCD have to be performed in each channel separately before the CCD is assembled.

\subsubsection{Calibration files}\label{sec:calib}

Bias frames were taken the same night as part of the observing program of A85. They consist of $15$ bias frames per CCD. However, there were only $2$ dark frames taken the same night. In order to derive a robust master dark for the images, we also downloaded the darks taken on adjacent nights; the 2014-09-22, 2014-09-23, and 2014-09-25, to a total of $10$ dark frames per CCD. The master bias and dark frames were created as the sigma-clipped ($3\sigma$) median for each of the channels of each of the CCDs. 

\subsubsection{Individual CCD processing}

In this step, we perform the processing and assembly of each CCD for each of the frames to produce a calibrated image. Each of the CCDs is processed independently. For each channel, we compute a median overscan using the corresponding overscan regions, and correct for overscan, dark and bias (as derived in Sec. \ref{sec:calib}). We also correct each channel for nonlinearity as done in the HSC pipeline \citep{Bosch2018} by applying a polynomial with coefficients determined per amplifier.
Before assembling the final CCD image, we applied the gains for each of the channels (provided in the headers). The final size of the assembled CCD is $2048\times4176$ pixels.

\subsubsection{Flat-field correction}\label{sec:flat}

An accurate estimation of the flat-field correction is crucial to achieving the goals of this study. Dome flats are not suitable for our goals due to inhomogeneities in the dome illumination that can result in gradients across the image \citep[e.g.,][]{Trujillo2016}. Consequently, our flat-field correction should be based on science exposures and, ideally, they should be the same science exposures used in this work. However, the images of the cluster are not appropriate for two reasons: 1) there are only 9 different exposures meaning that the resulting flats will be very noisy and 2) the offsets of the dithering pattern are not large enough for this purpose, so the galaxies of the cluster occupy roughly the same physical area of the CCD in all the exposures. The latter means that there will not be enough pixels to average in those regions in the resulting flats. 

To address this, we downloaded images of random fields from the HSC-SSP Wide survey taken on adjacent nights to the A85 observations in order to derive the most reliable flat-field correction possible. Using the SSP-Wide survey reduces the probability of an extended object in the same physical space of the CCD in all exposures. For the \g{} band the images were taken on 2014-10-01 (31), 2014-11-18 (9), 2014-11-25 (9), a total of 49 frames per CCD. The exposure times are $150$s per frame. 

For the \ib{} band, taking the images from the adjacent nights resulted in substructure remaining after the flat-field correction. This was found to be due to differences in the rotation angle of the instrument in the different set of images (see Appendix \ref{app:rot_flat} for more details). Therefore, the final images used were taken on 2014-03-27, 2014-09-17, 2015-01-22, 2015-07-11, 2015-07-20, 2014-09-22, 40 images per CCD in total. The exposure times are $200$s for each frame.

The assembled CCDs show a steep gradient across the detector that can cause detection algorithms to mistakenly detect and mask regions of the image that do not correspond to any source. To account for this, the construction of the flats was done in two steps. 

We first derived a rough flat or \emph{preflat}. These were derived by stacking the HSC-SSP science images using a median of the normalized images for each CCD, without any masking, to make a CCD flat. Each of the images that went into the flats was visually inspected to eliminate those presenting very bright, saturated stars and extended objects that might introduce errors in the derived flat-field frames.
First, we normalized each CCD image to one, using a region of $1000\times1000$ pixels located at the same position in the middle of each CCD. The \emph{preflats} were created as the sigma-clipped ($3\sigma$) median of the normalized images. Once these \emph{preflats} are derived, we use them to correct the assembled CCD images. We use these \emph{preflat}-corrected CCD images to build an object mask with \sextractor{} \citep{Bertin1996}. The settings used for the detection are optimized for faint object detection so to better mask faint sources. Again, for each CCD the masked and normalized images are combined to create the final flats.

Finally, each CCD is divided by the corresponding final flat. In Appendix \ref{app:flat_ex}, we show a region of our \ib{} band images where the improvement of using the flats with the same rotation as the science images can be seen.

\subsubsection{Astrometric calibration and frame assembly}\label{sec:mosaic}

Before combining the CCDs into frames, we need to refine the rough astrometry that the HSC camera provides. To do that, we use \scamp{} \citep{Bertin2006} to put the science images into a common astrometric solution. \scamp{} reads \sextractor{} catalogs and computes astrometric solutions for each individual CCD. The reference used is the stars of the SDSS DR9 catalogue \citep{Ahn2012} in our field of view. The number of stars used in each mosaic frame (40 CCDs) for our astrometric solution is typically around a couple of hundred.

After computing the accurate astrometry for each CCD in each frame, we need to make sure the CCDs are levelled before building the frame, i.e. all CCDs in the frame have the same sky counts. For each CCD, we run \sextractor{} again. We build a mask by using the segmentation map obtained, further expanding the detected objects by 10 pixels. In addition, we masked all bright sources in the CCDs. This includes bright stars to minimize the contamination of their extended halos, large galaxies and $\sim 700\arcsec$ in radius around the BCG. This constant correction is computed as the $3\sigma$-clipped median of the remaining pixels and subtracted from the respective CCDs.

After levelling each CCD, we use \swarp{} \citep{Bertin2002} to put together the $40$ CCDs from each exposure into single mosaic frames. \swarp{} resamples the CCDs putting them into a common grid using a LANCZOS3 interpolation function. The result is $9$ mosaic frames for both \g{} and \ib{} bands.

%
\begin{figure*}
\begin{center}
\includegraphics[scale=0.4]{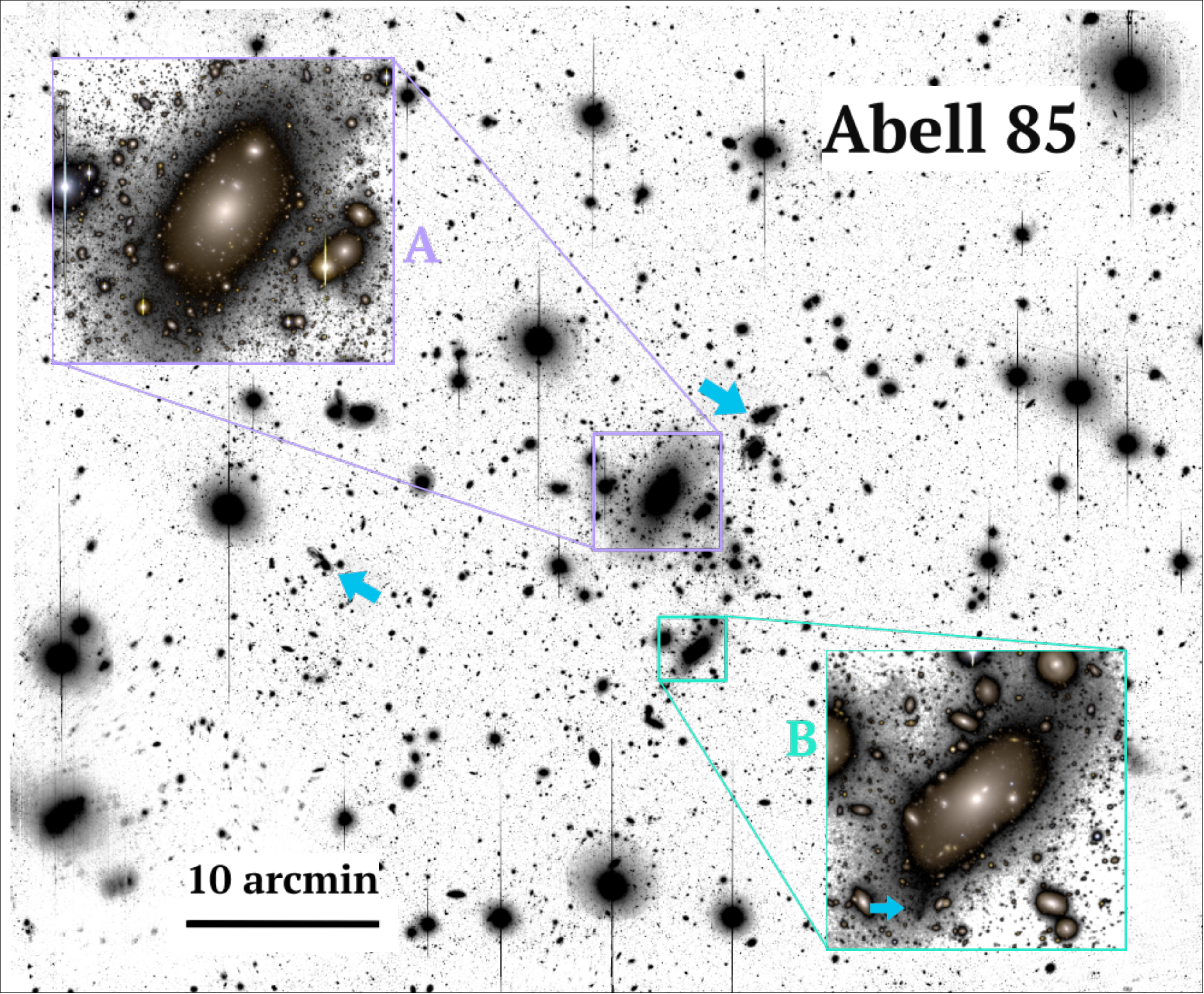}
\caption{Image of the cluster A85 in the \g-band. The area displayed is $52\arcmin \times 52\arcmin$ around the centre of the cluster (RA = 00h42m01.2s, DEC = -09d18m18.9s). Two regions of the cluster are highlighted. Zoom-in A ($390\arcsec \times 350\arcsec$, purple) shows an RGB image of the BCG of A85 where the ICL can be seen. Zoom-in B ($220\arcsec \times 200\arcsec$, light green) shows a satellite galaxy of the cluster. Tidal streams and other signs of interaction are easily seen when the images are zoomed-in on. Three of these features are marked with a blue arrow (one in the zoom-in B). The RGB images are a combination of the \g{}, and \ib{} bands whereas a black and white \g{} image is used for the main image. North is up and East is left. \label{fig:a85}}
\end{center}
\end{figure*}

\subsubsection{Sky subtraction}
Sky subtraction is one of the most important steps for reducing low surface brightness data as if done incorrectly it can introduce unwanted gradients or remove partially or entirely the object we want to study. The sky determination and subtraction is done for each of the mosaic frames individually before the final co-addition step. We first masked all sources in the individual mosaics using the segmentation maps provided by \sextractor{} and further dilated each object by 20 pixels to minimize contamination of the fainter regions of objects that are not included in \sextractor's segmentation map. Separately, we generously masked all bright sources (stars and galaxies) as well as the gaps between CCDs and created an additional mask to cover the centre of the cluster to avoid contamination of the outer parts of the BCG (as done in Sec. \ref{sec:mosaic} but now for the full mosaic). Once the mosaic is masked, we distributed $50,000$ boxes of $100\times 100$ pixels randomly through the image and computed the 3-$\sigma$ clipped median of the counts. We subtract this constant sky value from the respective mosaic.

In addition, we also fitted a first degree 2D polynomial to the masked mosaics. As the size of the mosaics is larger than the physical extent of the ICL in the images, this ensures the correction of any remaining gradients in the image while preserving the diffuse light in this cluster. This 2D polynomial is then subtracted from the entire mosaic.

\subsubsection{Image co-addition}
Once the science mosaics are sky-subtracted and in a common astrometric solution, we use \swarp{} to co-add the mosaics into a final image. \swarp{} projects the input images into the output frame and co-adds them in an optimum way. The method used for the geometric resampling is LANCZOS3. The final output is created as the median of the $9$ mosaic frames. Finally, we computed and subtracted a constant sky value from the final co-added images.

The final exposure times of the images are 1800s (30 mins) for the \g{} band and 2160s (36 mins) for the \ib{} band. The final \g{} band mosaic is shown in Fig. \ref{fig:a85}. The field of view is $52\arcmin \times 52\arcmin$. In Fig. \ref{fig:a85}, we also show RGB zoom-in images of two regions of the cluster. Region A shows a postage stamp of $390\arcsec \times 350\arcsec$ around the BCG of A85 (framed in purple) and region B shows a $220\arcsec \times 200\arcsec$ region around a massive galaxy belonging to one of the subclusters that are merging into A85 (\citealt{Ichinohe2015, Owers2017}; framed in green). The astrometric calibration is not accurate at the corners of our field of view, likely due to the lack of stars available to perform accurate astrometry there.

\subsubsection{Photometric calibration}

\begin{figure*}
\begin{center}
\includegraphics[width=0.75\textwidth]{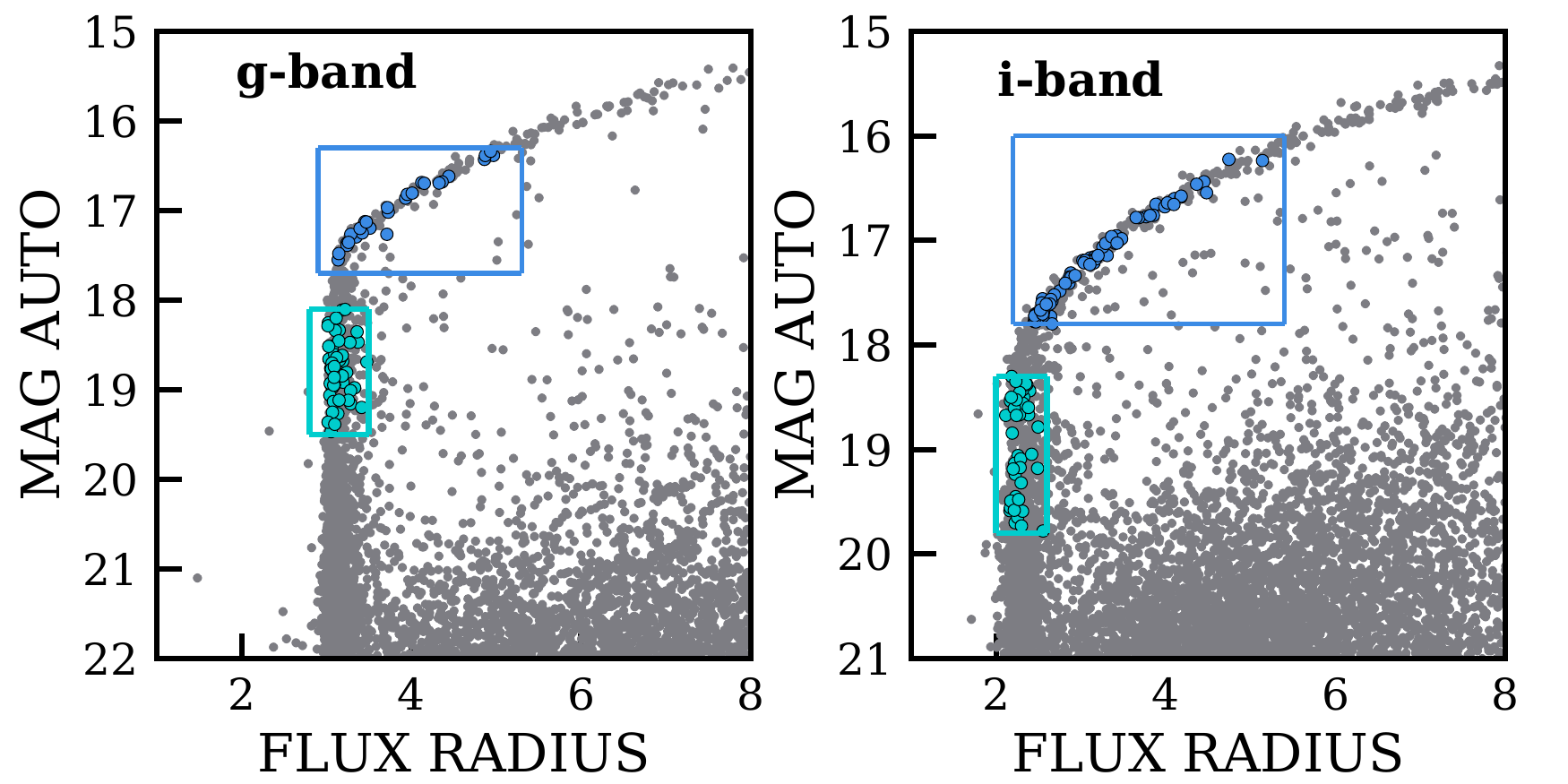}
\caption{Magnitude as a function of the half-light radius, in pixels, for all detected sources in the image of A85. The selection boxes for the stars used for the core (light green) and intermediate (blue) parts of the PSF are drawn and the selected stars are highlighted, for the \g{} (left panel) and \ib{} (right panel) bands.\label{fig:point_source}}
\end{center}

\end{figure*}

The photometric calibration of our images is based on the photometry of non-saturated stars in our field of view in common with the SDSS DR12 catalogue \citep{Alam2015}. For each band, we chose stars within the magnitude range (SDSS `psfMag') 18 to 21 mag, to avoid saturated stars in our mosaics, as seen in Fig. \ref{fig:point_source}, and very faint and noisy sources in SDSS. For our images, we used `MAG\_PETRO' which provides an estimate of the total flux of the star. We matched the SDSS DR12 photometric catalogue to ours, multiplying the frames by a factor to make the photometry in both catalogues equal. The typical number of stars that are used for photometric calibration within each individual mosaic image is $\sim 700$ stars. The average dispersion in the photometry for each band is $\sim0.1$ mag, for both the \g{} and \ib{} bands.

\subsection{Modeling and subtraction of stars} 

The careful modeling and removal of stars in deep images is now a common technique in low surface brightness science \citep[e.g., ][]{Slater2009, Trujillo2016, Roman2019}. This is important in order to minimize the contamination by light scattered by the stars, especially bright stars, in our photometry of the faint ICL. 

\subsubsection{Point spread function derivation}

A robust and extended characterization of the point spread function (PSF) of the image is crucial to remove the stars in the field of view, in particular bright stars close to the object of interest. For example, \citet{Uson1991b} showed that the total amount of diffuse light measured around the BCG of Abell 2029 would be in excess without removing nearby stars (their figure 5). 

In order to correct for this, we first construct the PSF of our images. Generally, to derive PSFs, we need to use stars with a wide range of brightnesses. The bright, saturated stars are used to characterize the outer parts of the PSF, or wings of the PSF, while fainter stars are used to characterize the core and intermediate parts. 

The bright stars in Fig. \ref{fig:a85} show asymmetries due to internal reflections in the telescope and the non-uniform illumination through it. These asymmetries become more significant further away from the centre of the camera. Given the limited amount of very bright stars in our image (N$\approx10$), we cannot build a PSF in every position of the camera. Luckily, the object of interest (BCG + ICL) is very close to the centre of the camera, therefore deriving a symmetric PSF to subtract nearby stars is a good approximation in this case.

\subsubsection{Core and intermediate part of the PSF}\label{sec:psf_inner}

In order to build the inner parts of the PSF, we followed a similar approach to the one in \psfex{} \citep{Bertin2011}. We first obtain a source catalog using \sextractor. The \sextractor{} catalog provides the half-light radius (`FLUX\_RADIUS') and the magnitude (`MAG\_AUTO') of the detected sources. It also provides the stellarity index `CLASS\_STAR' for discerning between stars and galaxies. A `CLASS\_STAR' close to 0 means that the object is very likely a galaxy, and 1 that it is a star. We select the objects of the catalog with `CLASS\_STAR' greater than 0.98. To minimize the asymmetries that can smear the structure of the PSF, we selected stars only in the inner $40\arcmin\times40\arcmin$ of the image. Their magnitude and half-light radius distribution is shown in Fig. \ref{fig:point_source}. We selected non-saturated stars (light green box) to derive the core, while brighter and saturated stars (blue box) are used to derive the intermediate parts of the PSF. 

We obtained the core and intermediate parts of the PSF by stacking the corresponding stars following these steps. First, we cut postage stamps around the selected stars of size $100$ and $500$ pixels$^2$ for the core and intermediate parts, respectively. In order to stack the stars, we need to accurately estimate their center. To do that, we need to mask all sources other than the selected star in the postage stamp. We use \sextractor's segmentation map for this. Then, we fitted a 2D Moffat model to the masked postage stamp of the star. Once the center of the Moffat is obtained, we re-centered the postage stamp. 

Second, we normalized each star by measuring a 1-pixel width ring at a radial distance of 15 pixels, avoiding the noisier outer parts (for the core stars) and the central saturated parts (for the intermediate stars). We also subtracted the sky around the stars in a 5 pixel-width ring at a radius of $13\arcsec$ for the core stars and $75\arcsec$ for the intermediate stars\footnote{These radii were defined to reach the background in each of the postage stamps, i.e., to not include flux from the star, at SNR$\sim1$.}. 

Finally, we stacked the normalized stars using a 3-$\sigma$ clipped median. The number of stars that were used for the stacking are 51 and 41, for the core, and 29 and 73, for the intermediate parts for the \g{} and \ib{} bands, respectively. 

\subsubsection{Outer parts of the PSF}\label{sec:psf_outer}

As discussed above, we want a model PSF that is extended enough that we can subtract the wings of stars close to the BCG + ICL system. However, in our field of view there are not enough bright stars to properly derive the outer parts of the PSF. This is also limited by the asymmetries that are more evident as we move away from the center of the image.  

For that reason, we selected a few very bright stars that are in our field of view and derived their radial profiles. The profiles of these stars look very similar despite the asymmetries, therefore we decided to use the radial profile of the closest bright star ($m_i\approx11$ mag, although saturated) to the center to build the outer part of the PSF. We adopted this methodology as a profile is more resistant to masking residuals or other artifacts that could bias the resulting PSF. It also means that this PSF will be symmetrical (i.e., we lose the spatial information). Note that the center parts of these very bright stars are strongly saturated causing bleeding in the detector, seen as spikes in Figure \ref{fig:a85}. We do not model these spikes.

We followed the same steps as for the core and intermediate parts. First, we cut a postage stamp of $2000$ pix $\times$ $2000$ pix around the star. We masked all sources that are not the selected stars using the segmentation map. In addition, to mask sources that are in the star's halo, we run \sextractor{} on an unsharp-masked image \citep{Sofue1993} of the postage stamp. The unsharp-masked image was obtained smoothing the stamp by a Gaussian with $\sigma = 30$ pix, which was then subtracted from the original. We combined both segmentation maps, from the original and the unsharp-masked image, to create the final mask.
We re-centered the postage stamp by fitting a Moffat2D model and shifting it to the new center given by the fit. In this case, the sky is subtracted at a distance of $325\arcsec$ to avoid contamination from the star flux (SNR$\sim1$). Then, we measured the radial profile of the star.

After deriving the radial profile of the star, we build the 2D outer PSF by assigning the value of each point of the profile to its corresponding radial distance ring around the centre. We then convolved the whole stamp with a $\sigma = 1$ pix Gaussian to smooth the abrupt changes at each given radius. This smoothing does not change the shape of the profile of the star.

Finally, we extend this outer part with a power-law in a similar way to \citet{MT18}. This last step is to minimise any sky subtraction issues in the outer parts of the star. We fit a power-law to the PSF image between $95\arcsec$ to $141\arcsec$, for the \g{} band, and $221\arcsec$ to $289\arcsec$, in the case of the \ib{} band\footnote{The bending in the profile of the \ib{} band at $200\arcsec$ is seen in the bright stars' postage stamps as well as in their profiles and is not a consequence of the sky subtraction.}. This power-law fit was used to extrapolate the outer regions of the PSF to a radius of $420\arcsec$.

\begin{figure}
\begin{center}
\includegraphics[width=0.45\textwidth]{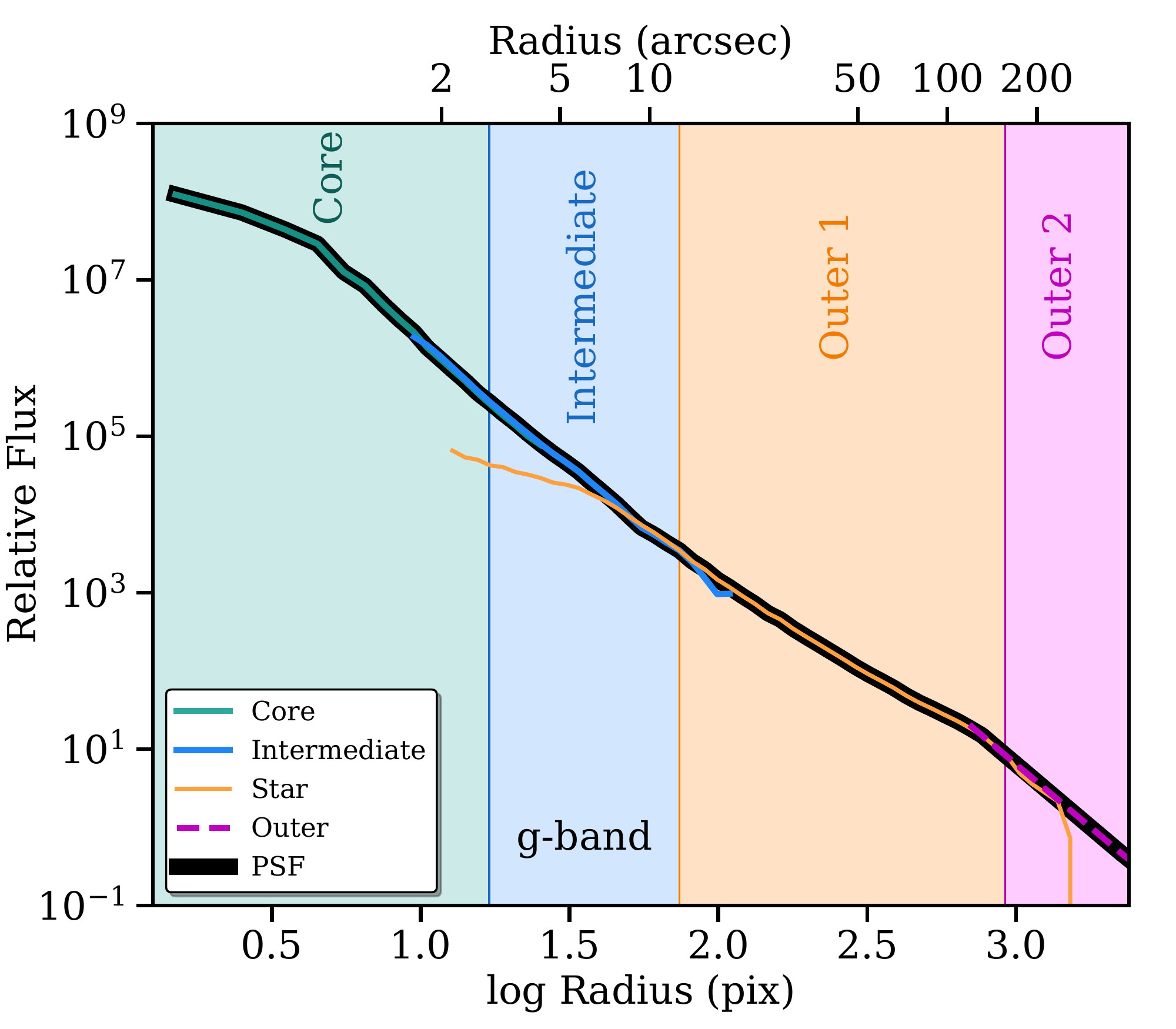}
\caption{Radial PSF profile in the HSC \g{} band in black. The different shaded regions correspond to the four different parts derived in Sec. \ref{sec:psf_inner} and \ref{sec:psf_outer} from which the final PSF was constructed. The colored lines are the individual radial profiles of the four different parts. The vertical lines that divide the shaded regions indicate the radii at which these different parts were joined.
\label{fig:psf_g}}
\end{center}

\end{figure}

\begin{figure}
\begin{center}
\includegraphics[width=0.45\textwidth]{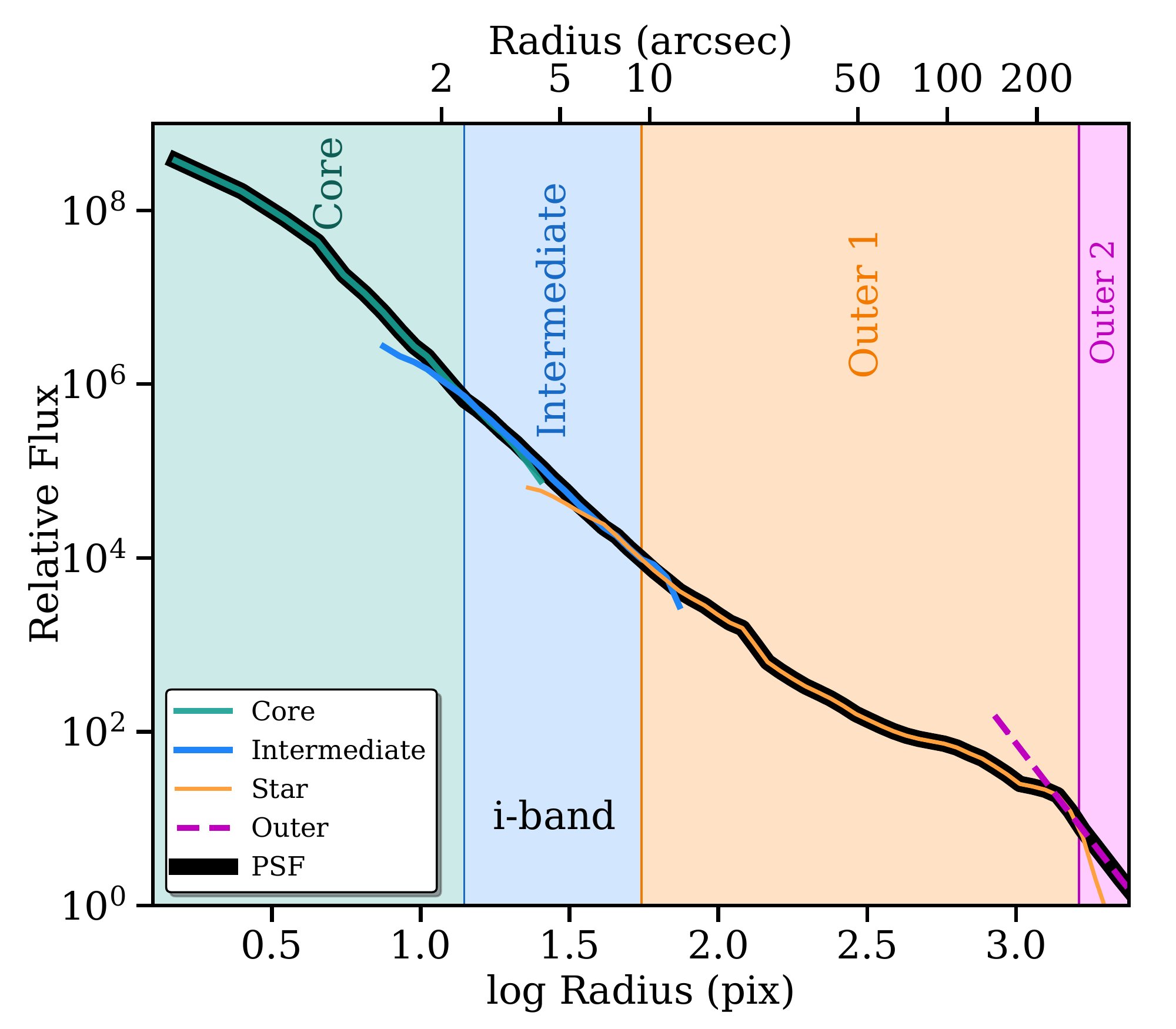}
\caption{Same as Fig. \ref{fig:psf_g} but for the HSC \ib{} band.\label{fig:psf_i}}
\end{center}

\end{figure}

\subsubsection{Connecting the different parts of the PSF}\label{sec:psf_final}

Once we derived the four different parts described above, we constructed our final PSF. We follow a similar approach to \citet{Infante-Sainz2020}. We use the radial profile of the bright star derived above as a reference for the connection and multiply the other profiles by a factor so they match the profile of the bright star at a given radius. The radius at which these connections are made change depending on the band. 
Fig. \ref{fig:psf_g} and Fig. \ref{fig:psf_i} show the final PSF profiles (black thick line) for the \g{} and \ib{} bands, respectively. The shaded regions indicate the four different parts used to construct the final PSF derived in Sec. \ref{sec:psf_inner} and \ref{sec:psf_outer}. The profile of the bright star, which was used for building the outer part of the PSF is labelled as \emph{Outer 1} in orange. The power-law extrapolation to the bright star profile is \emph{Outer 2}, in magenta. The core and intermediate parts are in teal and blue, respectively. We also show the different individual profiles used to construct the final PSF, in their respective colors. The radii where the connections were made for each band and each of the different parts are indicated by the vertical lines in the plots, in teal (connection between core and intermediate part), orange (between intermediate and the bright star profile) and magenta (between the bright star profile and the power-law extension). The total flux of the final PSFs (\g{} and \ib{}) is normalized to 1.

\subsubsection{Star subtraction}

\begin{figure*}
\begin{center}
\includegraphics[width=\textwidth]{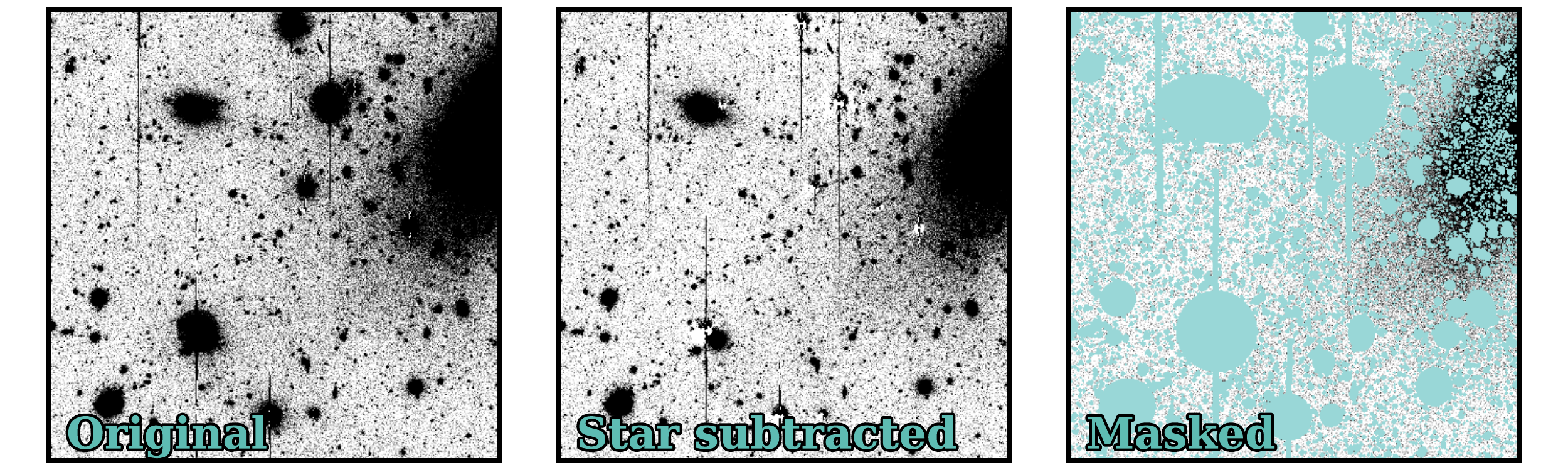}
\caption{Example of the star subtraction and masking process in a $500\arcsec\times500\arcsec$ region of the image of A85 for the \ib{} band. The images shown have the same contrast. \label{fig:ex_mask}}
\end{center}
\end{figure*}

To subtract the stars in our images, we follow similar steps to \citet{Roman2019}. We started by building a catalogue of the positions of visually-selected bright stars. There are two key aspects when fitting these stars: to obtain an accurate centre of the star and to perform the flux calibration. We produced postage stamps for each of the stars of $500\times500$ pixels. Then, we masked all sources that are not the central star to avoid contamination that could affect the flux calibration and centering. 
This masking is done in two steps: 1) a first run to detect all sources with and 2) a second run where the detection image is an unsharp-masked image, with a Gaussian smoothing of $\sigma = 20$ pix. This second step allows us to mask sources that are covered by the halo of the star and not properly detected in the first run. In both cases, the detection was done with \sextractor.

To accurately center the star, we calculated the centroids for each star by fitting a 2D Gaussian to the 2D flux distribution of the star using \texttt{centroid\_sources} in \texttt{photutils}. \texttt{centroid\_sources} allows us to define a box for fitting the Gaussian, useful in cases where the centre of the star is strongly saturated. 

Once the star is masked and centered, we performed the flux calibration. We first derived radial profiles for both star and the PSF. By using the profiles rather than the stamps we are minimizing contamination due to masking residuals or other artifacts. To fit each star, we selected a range in radius for the calibration. The radial range is from $0.1$ times the saturation level of the image to $4$ times the value of the background of each postage stamp. This background was calculated as the standard deviation of the postage stamp with all the sources masked (including the star).  

We scaled the PSF profile to match the star profile, using ratio between star and PSF values derived from the profiles. Once the PSF is centered and calibrated we subtracted it from the image. We repeated the same process for each of the stars in the catalogue for both \g{} and \ib{} bands. Fig. \ref{fig:ex_mask} shows a region of our image of A85 in the \ib{} band. The original image is seen in the left panel while the middle panel shows the same region with the stars subtracted.

As mentioned above, the stars in HSC images show asymmetries that become more evident further away from the center of the image. However, we have built a symmetric PSF. As the object of interest, the BCG, is centered in the image, nearby stars that could affect our photometry are not going to present significant asymmetries. However, we note that this is a potential source of error for this study. 

\subsection{Masking}\label{sec:masking}

The study of the ICL in clusters of galaxies requires a very careful masking of foreground and background sources to reduce contamination that can affect the determination of the color of this light. In the case of deep images, this masking must be optimized not only for faint and small background objects but also for those that are closer and large. 

As a single setup for the detection and masking of both types of sources is unfeasible, we used a two-step approach like \citet{MT18}; a ``hot+cold'' mode \citep[e.g.,][]{Rix2004}. The ``cold'' mode will detect the extended bright galaxies from the cluster while the ``hot'' mode is optimized to detect the faint and small sources. We use this approach on a deep combined $g+i$ image, after star subtraction. In the case of the ``hot'' mode, we unsharp-masked the original image, to enhance the contrast, particularly in the central parts of the BCG. To create the unsharp-masked image, we convolved the image with a box filter with a side of $25$ pixels and then subtracted it from the original. The threshold for detection is $1.1\sigma$ above the background.

The ``cold'' mask was further expanded 10 pixels while the ``hot'' was expanded 5 pixels. Both masks were combined to create the final mask for our images. Before this, we unmasked the BCG on the ``cold'' mask. The bleeding spikes were manually masked as well as the residuals from the subtraction of stars and their asymmetries. 

We created two masks for the cluster. In the first mask, all the objects of the image are masked except for the members of the cluster contained in our field of view and the diffuse ICL. For that, we use the spectroscopic membership information obtained in \citet{Owers2017}. The morphological information obtained from \sextractor{}'s ``cold'' mask run is used to unmask the members of the cluster.

For the second mask, all the objects are masked except for the BCG and ICL. As \sextractor{} does a poor job detecting low surface brightness outskirts, we manually extended the masks for the remaining objects after visual inspection.

The final mask was again visually inspected to manually mask any remaining light that was missed by the process described above. In the right panel of Fig. \ref{fig:ex_mask}, we show an example of the mask in one region of our image. 

\subsection{Surface brightness limits}

Our goal is to study the low surface brightness features in A85 down to the faintest surface brightness possible. For this reason, we need to know how deep our images are by estimating the surface brightness limits that they reach.
To obtain these limits, we calculated the r.m.s of the final masked images by randomly placing $20000$ boxes of $10\times10$ arcsec$^2$ ($\sim10\times 10$ kpc$^2$) across the images. In this case, we also masked the BCG and ICL by adding an ellipse of semi-major axis of $672\arcsec$ centered in the image. 

The $3\sigma$ surface brightness limits are: $\mu_{g}^{limit}$(3$\sigma$, $10\arcsec\times10\arcsec$) = 30.9 mag/arcsec$^2$, and $\mu_{i}^{limit}$(3$\sigma$, $10\arcsec\times10\arcsec$) = 29.7 mag/arcsec$^2$. These limits are calculated following Appendix A in \citet{Roman2019}.

\section{The intracluster light of A85}

\subsection{Radial surface brightness profiles} \label{sec:profiles}

\begin{figure*}
\begin{center}
\includegraphics[width=0.99\textwidth]{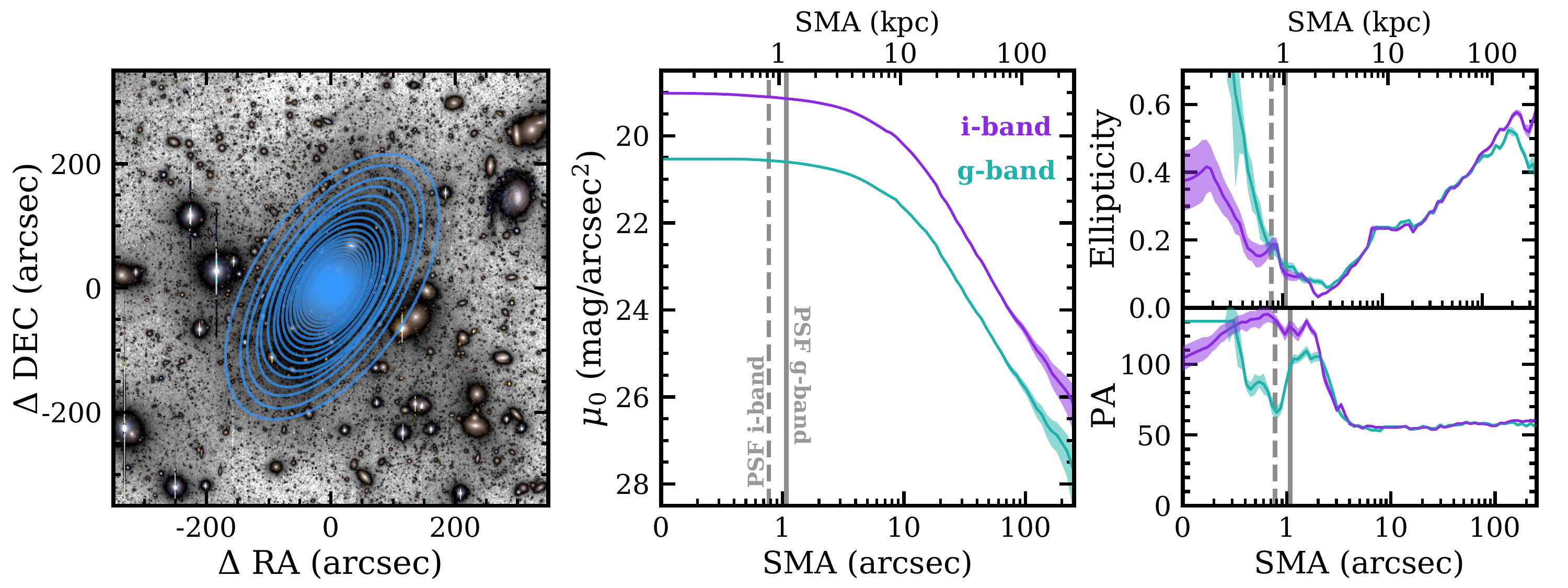}
\caption{The left panel shows the inner $700\arcsec\times 700\arcsec$ of the A85 image with the isophotes from \texttt{ellipse}. The middle panel presents the surface brightness profiles as a function of the semi-major axis for the BCG + ICL of A85 to a radius of $250\arcsec$ (258 kpc), for the \g{} (green) and \ib{} (purple) bands. The profiles are $k$-corrected and corrected for the extinction of the Milky Way and surface brightness dimming. The right panel shows the ellipticity (top) and position angle (PA, bottom) profiles as a function of the semi-major axis for the \g{} and \ib{} bands. The vertical gray lines indicate the FWHM of point sources in the \g{} (solid) and \ib{} (dashed) images. 
\label{fig:profiles}}
\end{center}
\end{figure*}

The goal of this paper is to study the diffuse light in HSC images of A85. To that end, we derived the radial profiles for the \g{} and \ib{} bands using the software \texttt{ellipse} in IRAF. \texttt{ellipse} fits elliptical isophotes to the 2-D images of galaxies using the method described in \citet{Jedrzejewski1987}. It provides the mean intensity, ellipticity, position angle and harmonic amplitudes (deviations from perfect ellipticity) for each fitted isophote. By deriving the 1-D profiles this way, we are not assuming any particular model or models to describe the BCG+ICL, as they might be sensitive to the choice of the particular model and prone to degeneracies between the different parameters.

\texttt{ellipse} was run on the star-subtracted, masked images. We first run the task allowing all parameters to vary freely. In the second run, we fixed the centers to the median centers of the isophotes returned by \texttt{ellipse} in the first iteration.We adopted the median setup in \texttt{ellipse}. The surface brightness profiles reach a signal-to-noise ratio of $2.8$ ($3.0$) at $27.1$ ($25.7$) mag/arcsec$^2$ in \g{} (\ib), which corresponds to a radius of $200\arcsec$ or $213$ kpc. 
Fig. \ref{fig:profiles} shows the output of \texttt{ellipse} for A85. The left panel shows the $700\arcsec\times700\arcsec$ region around the BCG\footnote{Known as Holm 15A.} with the fitted ellipses. The 1-D radial surface brightness profiles as a function of semi-major axis (SMA) for the \g{} (green) and \ib{} (purple) bands are shown in the middle panel, up to $250\arcsec$. These surface brightness profiles are corrected for the absorption of our galaxy \citep[E(B-V) $= 0.034$;][]{Schlafly2011} and surface brightness dimming. The profiles are also $k$-corrected \citep{Chilingarian2010, Chilingarian2012}. The shaded regions represent the errors of the profiles computed as the r.m.s scatter of each isophote.

The vertical gray lines in all the panels indicate the full-width-at-half-maximum (FWHM) of the \g{} (solid; $1\farcs07$) and \ib{} (dashed; $0\farcs78$) bands. The FWHM of each image is given by twice the average `FLUX\_RADIUS' (the half-light radius) of stars obtained from \sextractor{} (see Fig. \ref{fig:point_source}). These lines define the regions where the isophotal fits are not reliable. The right panel shows the ellipticity (top) and position angle (PA; bottom) with SMA for both bands.

The surface brightness profiles derived here show a flattening in the central regions of the BCG ($\lesssim 10\arcsec$, $11$ kpc). This flattening in the inner $\sim10\arcsec$ has already been reported \citep[e.g.,][]{Lopez-Cruz2014, Madrid2016}. In fact, the BCG of A85 is known to host one of the largest cores measured to date \citep{Lopez-Cruz2014}.

Beyond the core, the surface brightness radial profiles roughly follow a \citet{Sersic1968} profile. However, in the middle panel of Fig. \ref{fig:profiles}, there appears to be a break in the profile at a radius of $\sim 70\arcsec$ ($\sim 75$ kpc), where the profiles become shallower.

\begin{figure}
\begin{center}
\includegraphics[width=0.48\textwidth]{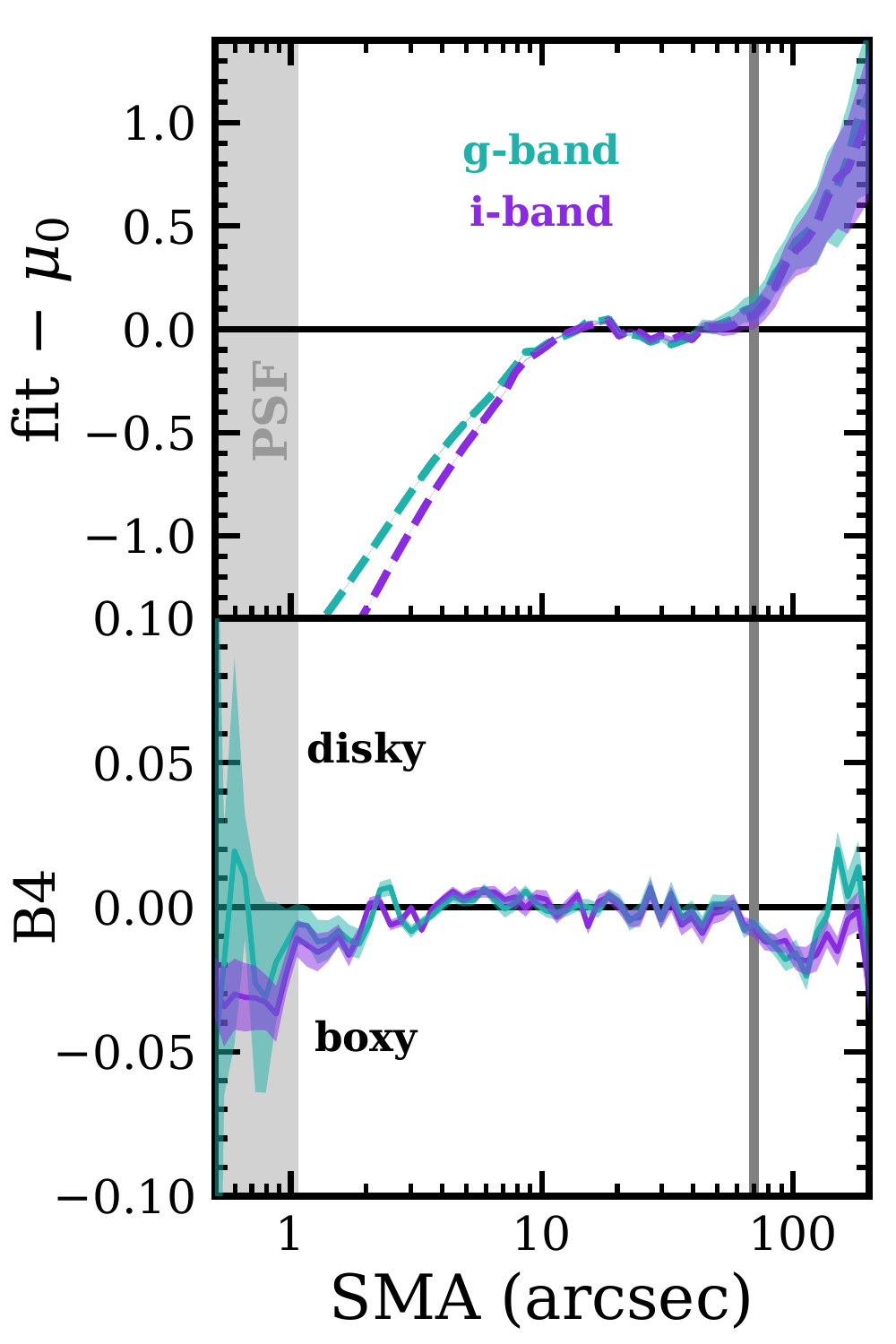}
\caption{The top panel shows the residuals from a S\'ersic fit to the surface brightness radial profiles in the \g{} (green) and \ib{} (purple) bands. The bottom panel shows the B4 coefficient (4$th$ Fourier harmonic) as a function of semi-major axis for both bands. The vertical grey line at $70\arcsec$ ($\sim75$ kpc), tentatively marks the radius where an extra component starts to dominate and the isophotes become boxier. \label{fig:fit_b4}}
\end{center}
\end{figure}

In order to explore whether there is a break in the surface brightness radial profiles, we fit a single S\'ersic profile to both bands, excluding the inner $10\arcsec$. These fits are performed using a least squares fitting method as suggested in \citet{Seigar2007}. The best fit to the profiles are shown in Fig. \ref{fig:fit_sersic} and the parameters are listed in Appendix \ref{app:sersics}. We show the residuals of subtracting the best S\'ersic fit from the surface brightness profiles in the top panel of Fig. \ref{fig:fit_b4}. The figure shows that at a radius of $\sim70\arcsec$ ($\sim75$ kpc) there is an excess of light with respect to the S\'ersic fit.\footnote{Note that the goal of this fit is to locate the break, not to describe the light profile.} This indicates that there is an extra component over the galaxy; the ICL. The position of the break found here is consistent with \citet{Zibetti2005}, where a similar flattening is found at a radius of $\sim 80$ kpc, in their stacked profiles of multiple clusters.

The ellipticity of the diffuse light of A85 increases with radius up to a value of $\epsilon \sim 0.55$ for both bands at a radius of $\sim 200\arcsec$ ($\sim213$ kpc), as shown in the top right panel of Fig. \ref{fig:profiles}. \citet{Kluge2020} also observed an increase in ellipticity for A85. However, at a radius of $\sim250$ kpc, their ellipticity profile drops sharply to a value of $0.1$. We do not see any evidence of such a decrease in our profiles\footnote{Our ellipticity profile remains constant at $\sim 0.5$ to a radius of $320$ kpc, although the signal-to-noise at that radius is $\lesssim 1$}. In contrast, the PA does not show any significant change with radius.

Departures from perfect elliptical isophotes can be described as Fourier harmonic perturbations \citep{Jedrzejewski1987}. The coefficients of these harmonic series carry physical meaning. For example, B4, the 4$th$ Fourier amplitude, indicates the boxyness/diskyness of the isophotes. In the bottom panel of Fig. \ref{fig:fit_b4}, we show the B4 coefficient as a function of SMA. The radius where the break of the surface brightness profile is located, $70\arcsec$, also corresponds to where the B4 becomes negative, i.e. the ellipses start showing a boxy shape. This radius is indicated in both panels of Fig. \ref{fig:fit_b4} by a gray vertical line. This is a confirmation of the boxyness visible in the outer parts of the BCG (inset A in Fig. \ref{fig:a85}). Boxyness has been found to be related to galaxy interactions \citep[e.g.,][]{Nieto1989}.

\subsection{Color profile of the BCG+ICL}\label{sec:color}

\begin{figure}
\begin{center}
\includegraphics[width=0.47\textwidth]{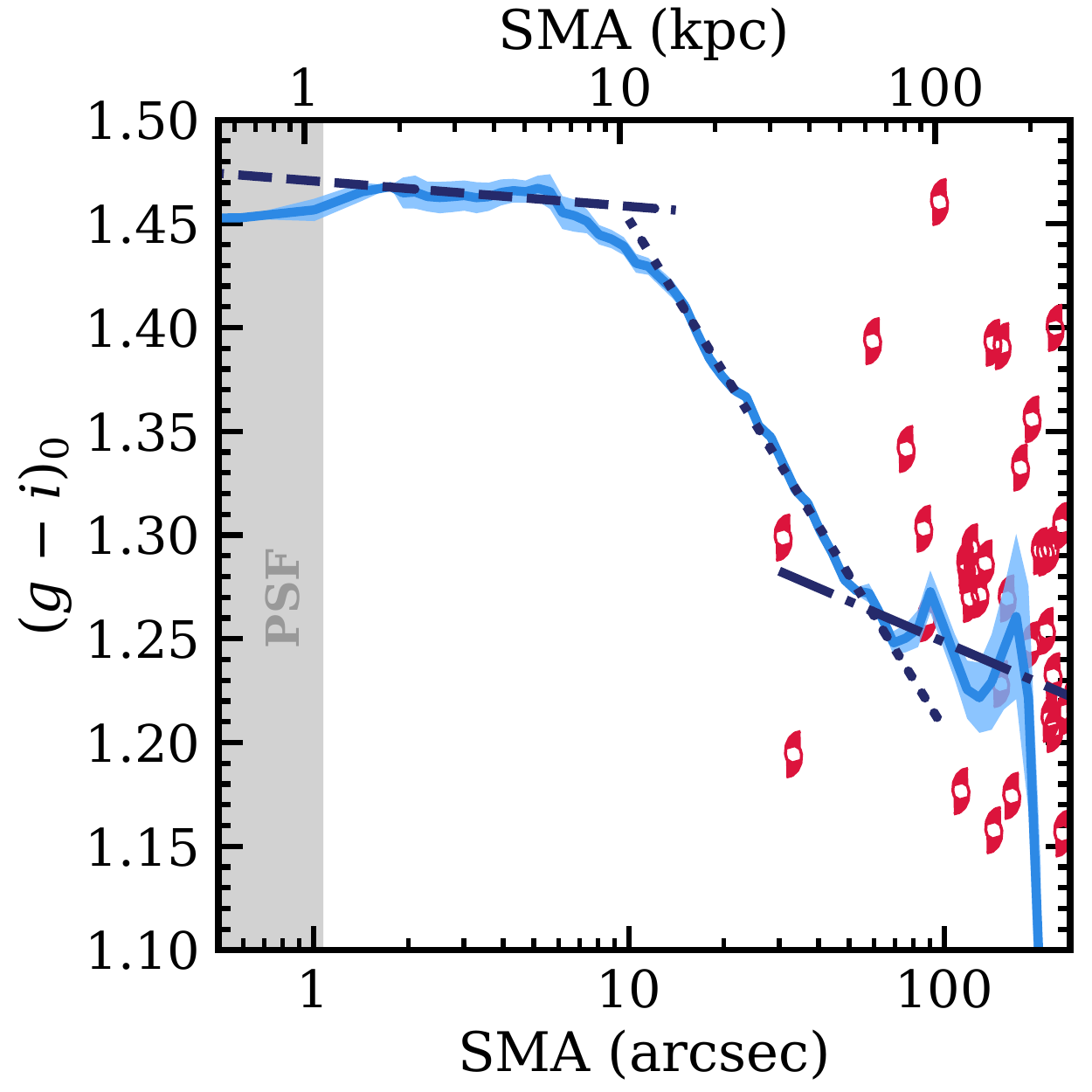}
\caption{The $g-i$ color profile of BCG + ICL of A85 in blue. The errors are indicated by the light blue shaded area. The red spirals indicate the average color of member galaxies of the cluster as derived in \citet{Owers2017}. The color profile presents three different regions: a flattening at $<10\arcsec$, a color gradient from $10\arcsec$ to $75\arcsec$ and a region from $70\arcsec$ to $200\arcsec$ where the gradient shallows. The linear fits to the color profiles for each region are shown as the dark blue lines: dashed for $<10\arcsec$, dotted for $10\arcsec$ to $75\arcsec$ and dash-dotted for $75\arcsec$ to $242\arcsec$. \label{fig:color}}
\end{center}

\end{figure}

Radial color gradients provide valuable constraints in the formation processes of galaxies and, consequently, the BCG and ICL \citep[e.g.,][]{MT14, MT18}. The radial color profile was measured in $55$ logarithmic spaced bins from 0 to $200\arcsec$. The distance to each pixel on the images is computed as the elliptical distance to the BCG, where the morphological parameters (ellipticity and PA) are the median values from the \texttt{ellipse} isophotes excluding the inner $10\arcsec$: $0.37$ for the ellipticity and $56\deg$ for the PA. For each radial bin, the surface brightness in each band was obtained by averaging the pixel values. The errors are drawn from jackknife resampling, i.e. repeating the photometry in a sub-sample of the data for each bin. The number of sub-samples per bin was 100.
Fig. \ref{fig:color} shows the $g-i$ color profile for the BCG+ICL of A85 down to $200\arcsec$ ($213$ kpc; light blue line). The color profile is $k$-corrected and corrected for the extinction of the Galaxy. The error in the color profile, represented as the light blue area, is the sum of the errors in the individual surface brightness radial profiles. We have also plotted the $g-i$ color of the satellite galaxies in the cluster as reported by \citet{Owers2017}.

The color profile of the BCG + ICL shows three distinct regions: i) a flat region out to $10\arcsec$ indicative of the core of the galaxy, ii) a negative color gradient from $10$ to $\sim70\arcsec$ and iii) a region from $\sim70\arcsec$ to $\sim200\arcsec$ where the color gradient of the diffuse light becomes shallower. To see if there is a difference, we calculated the gradients of each region as a linear fit to the color profile $g-i$ vs. log R ($\Delta gi$). The fits are shown in Fig. \ref{fig:color} as the dark blue lines. The gradients for the different regions are: i)  $-0.01\pm 0.01$ (dashed line), ii) $-0.24 \pm 0.01$ (dotted line) and iii) $-0.06 \pm 0.04$ (dash-dotted line).

The flat color profile at SMA $<10\arcsec$ ($<11$ kpc) coincides with the size of the core of the galaxy as seen by \citet{Lopez-Cruz2014}. This is consistent with a mixing of the stellar populations in the centre of the galaxy.

The region between $10\arcsec$ to $\sim75\arcsec$ ($11$ to $\sim80$ kpc) presents a negative gradient in $g-i$ color from $1.45$ to $\sim1.25$ ($\Delta gi = -0.24 \pm 0.01$). It is well known that massive early-type galaxies have negative optical color gradients indicating gradients in their stellar populations, generally metallicity \citep[e.g.,][]{Peletier1990, Davies1993, LaBarbera2012, Huang2018, Santucci2020}. 

Beyond $\sim75\arcsec$ ($\sim80$) kpc, the color profile becomes significantly shallower ($\Delta gi = -0.06 \pm 0.04$) with a median color of $g-i = 1.25$. The observed behaviour of the color profile of A85 is consistent with the color profile in \citet{Zibetti2005} \citep[also, ][]{Coccato2010, Montes2014a, Spavone2020}. \citet{Zibetti2005} explored the $g-r$ color profile of stacked clusters in SDSS. Their color profile also shows a gradient down to $\sim80$ kpc where it shallows.

\begin{figure*}
\begin{center}
\includegraphics[width=0.9\textwidth]{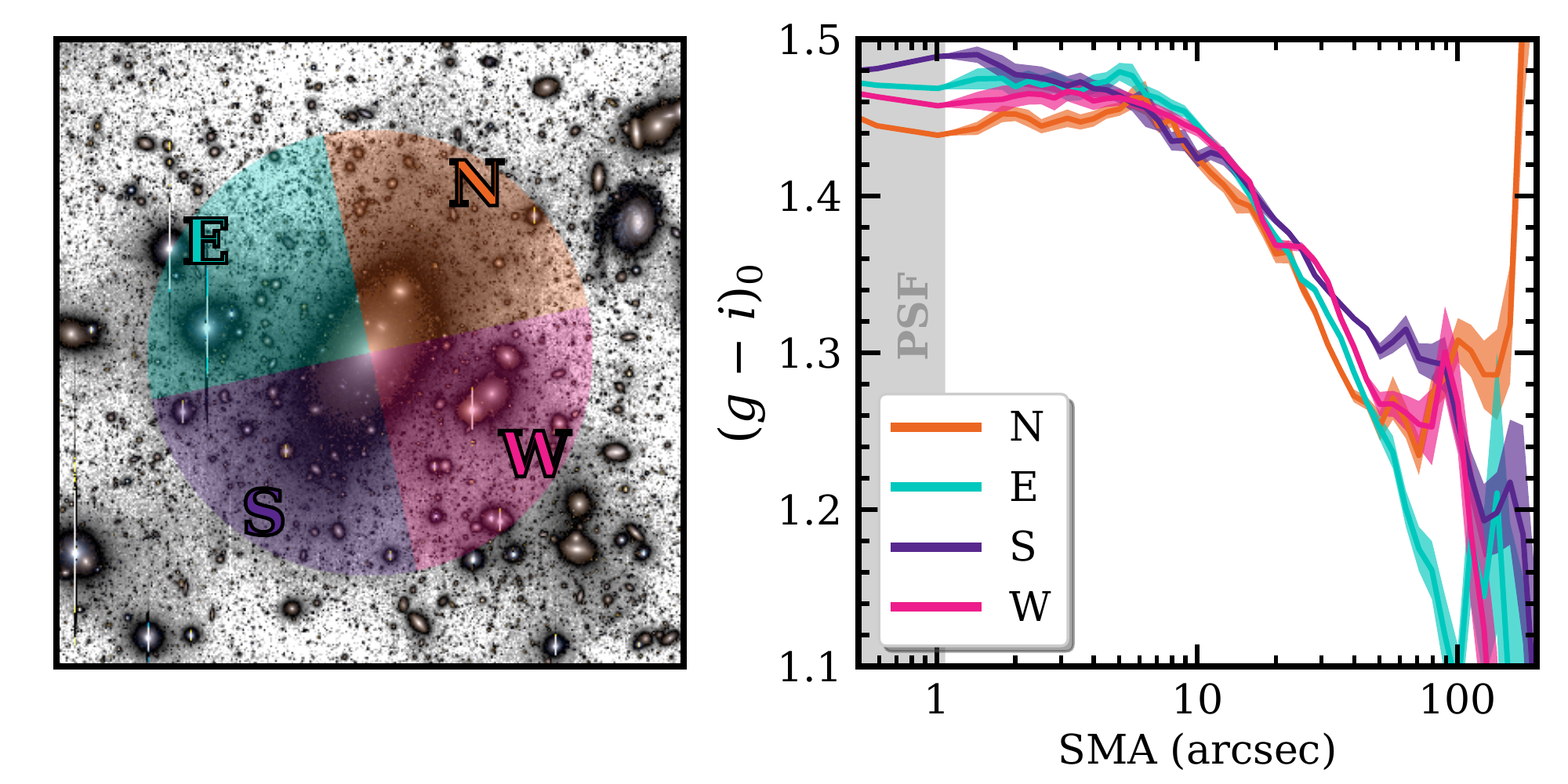}
\caption{The left panel shows the inner $700\arcsec\times700\arcsec$ of A85 where the different sections are shaded in different colors: North (N, orange), East (E, green), South (S, purple) and West (W, magenta). In the right panel we show the corresponding $g-i$ color profiles to a radius of $200\arcsec$. The Northern, Southern and Western profiles flatten at different radius possibly indicating the presence of accreted material at those distances.  \label{fig:quarters}}
\end{center}
\end{figure*}

There are some nearby bright stars both East and West of the BCG. Inaccuracies in the star subtraction process could bias the colors that we obtain, particularly the colors of the faintest regions of the ICL. In order to assess that potential issue, we derive the color profiles in 4 different directions of the BCG: North, East, South and West. 

The four different profiles were derived by masking the BCG+ICL except for $90\deg$-wide sections as shown in the left panel of Fig. \ref{fig:quarters}, labelled as North (orange, N), East (green, E), South (purple, S) and West (magenta, W). The profiles are derived in the same way as the overall color profile.
The right panel in Fig. \ref{fig:quarters} shows the color profiles color-coded by their respective section. The color profiles behave similarly up to $\sim 50\arcsec$, where the Southern profile flattens (purple line, $g-i\approx1.3$) to a radius of $\sim100\arcsec$ ($107$ kpc). Similarly, the Northern (orange) and Western (magenta) profiles show flattening, and even reddening (North profile), between $80\arcsec$ to $130\arcsec$. 

While for the Southern profile there is not a clear origin for the observed flattening, the shape of the Northern profile could be affected by the presence of a large satellite (at a projected radius of $\sim 75\arcsec$, zoomed-in in Fig. \ref{fig:res}). This is also the case for the Western profile as there are some galaxies at $\sim145\arcsec$. We will discuss this in detail in the following Section.

Given that the closest bright stars are only located East and West of the BCG, these color profiles confirm that the change in gradient is not caused by the presence of these stars but rather caused by the presence of diffuse light associated with ongoing mergers.

\subsection{Fraction of light in the ICL}

Studying the amount of light in the ICL can provide information on the efficiency of the interactions that form the ICL. This is given by the ICL fraction, defined as the ratio between the ICL and the total (BCG + galaxies + ICL) flux or luminosity of the cluster.
This ICL fraction is an ill-defined quantity in photometric-only studies as separating between BCG and ICL is not obvious. To overcome this problem, astronomers have been using different ways of defining the ICL component in deep photometry. In the following, we describe two of the most widely used definitions. We derived the ICL fraction for A85 using both of them, for ease of comparison with other studies. 

\subsubsection{ICL fraction from surface brightness cuts}

The most widely used definition is to apply a cut in surface brightness and assume that the light fainter than a certain surface brightness limit is the ICL \citep[typically $\mu_V > 26.5$ mag/arcsec$^2$, e.g.,][]{Feldmeier2004, Rudick2011}. To derive the ICL fraction for the \g{} and \ib{} bands, we followed similar steps to \citet{MT18}. First, we applied the mask where all the members of the cluster are unmasked, derived in Sec. \ref{sec:masking}, to each of the images. In each of the bands, we summed all the pixels fainter than a given ICL threshold. The fractions given have a fainter limit of $\mu< 29.5$ mag/arcsec$^2$ in order to minimize the contamination from inhomogeinities in the background. The ICL fractions are derived applying 3 different surface brightness cuts: $\mu>$ 26, 26.5 and 27 mag/arcsec$^2$. We provide the ICL fractions for both the \g{} and \ib{} bands in Table \ref{tab:fractions}..

The ICL fractions calculated this way account not only for the diffuse light associated with the BCG but also with other galaxies in the cluster. Note that defining the ICL this way means that the measured fractions are a lower limit of the true value; we are missing light in both the brighter (e.g., in projection) and fainter limit.

\begin{deluxetable*}{lcccc}[t]
\tabcolsep=0.4cm
\tablecaption{\label{tab:fractions}
	ICL fraction ($\%$) for A85}
\tablehead{ \multicolumn{5}{c}{Surface brightness cuts} \\
& $26<\mu<29.5$ &  $26.5<\mu<29.5$ & $27<\mu<29.5$ & $27.5<\mu<29.5$ \\
\multicolumn{5}{c}{[mag/arcsec$^2$]}}
\startdata
$f_{\mathrm{ICL}}(g)$ & $8.8\pm0.5$ & $6.2\pm0.7$ & $4.0\pm0.9$ & $2.4\pm0.9$ \\
$f_{\mathrm{ICL}}(i)$ & $3.1\pm0.7$ & $1.9\pm0.7$ & $1.1\pm0.7$ & $0.6\pm0.7$ \\
\hline
\hline
\multicolumn{5}{c}{2D fit}\\
& \multicolumn{2}{c}{\g}  & \multicolumn{2}{c}{\ib}\\
\hline 
$f_{\mathrm{ICL}}$ &\multicolumn{2}{c}{$11.0\pm1.0$} & \multicolumn{2}{c}{$11.5\pm1.0$} \\
$f_{\mathrm{BCG+ICL}}$ &\multicolumn{2}{c}{$16.7\pm2.0$} &\multicolumn{2}{c}{$18.0\pm2.0$} \\
$f_{\mathrm{ICL}/\mathrm{BCG+ICL}}$ &\multicolumn{2}{c}{$66.1\pm2.2$}&\multicolumn{2}{c}{$63.7\pm2.2$}\\
\hline
\enddata
\end{deluxetable*}

\subsubsection{ICL fraction assuming a functional form}
Despite its simplicity, one of the limitations of the above definition is that it does not account for the amount of ICL in projection on top of the BCG. Another common approach is using functional forms to describe both BCG and ICL \citep[e.g.,][to name a few]{Gonzalez2005, Seigar2007, Spavone2018}. In our case, we use GALFIT \citep{Peng2002} to simultaneously fit two two-dimensional S\'ersic profiles: one to describe the BCG and one for the ICL. The parameters for the two fitted S\'ersic components are given in Table \ref{tab:sersic2d} in Appendix \ref{app:sersics}. Although the fits seem to describe well the overall profile, they are not able to reproduce the inner core of the galaxy (as in the case of the single S\'ersic fit, Fig. \ref{app:sersics}).
Contrary to the single S\'ersic fit performed in Sec. \ref{sec:profiles}, we now find that the inner component is an exponential, similar to the outer component ($n_1 \sim 1$ and $n_2 \sim 2.15)$. This difference between the single and double S\'ersic fits is probably caused by the single component trying to fit the outer parts of the BCG+ICL profile.

As expected from the \texttt{ellipse} 1-D profiles in Sec. \ref{sec:profiles}, the more extended component (the ICL) has a higher ellipticity than the inner component \citep[the BCG, see also][]{Kluge2020b}. However, the PA in both models are not significantly different ($\Delta \mathrm{PA} \sim 4^{\circ}$).

\begin{figure}
\begin{center}
\includegraphics[width=0.45\textwidth]{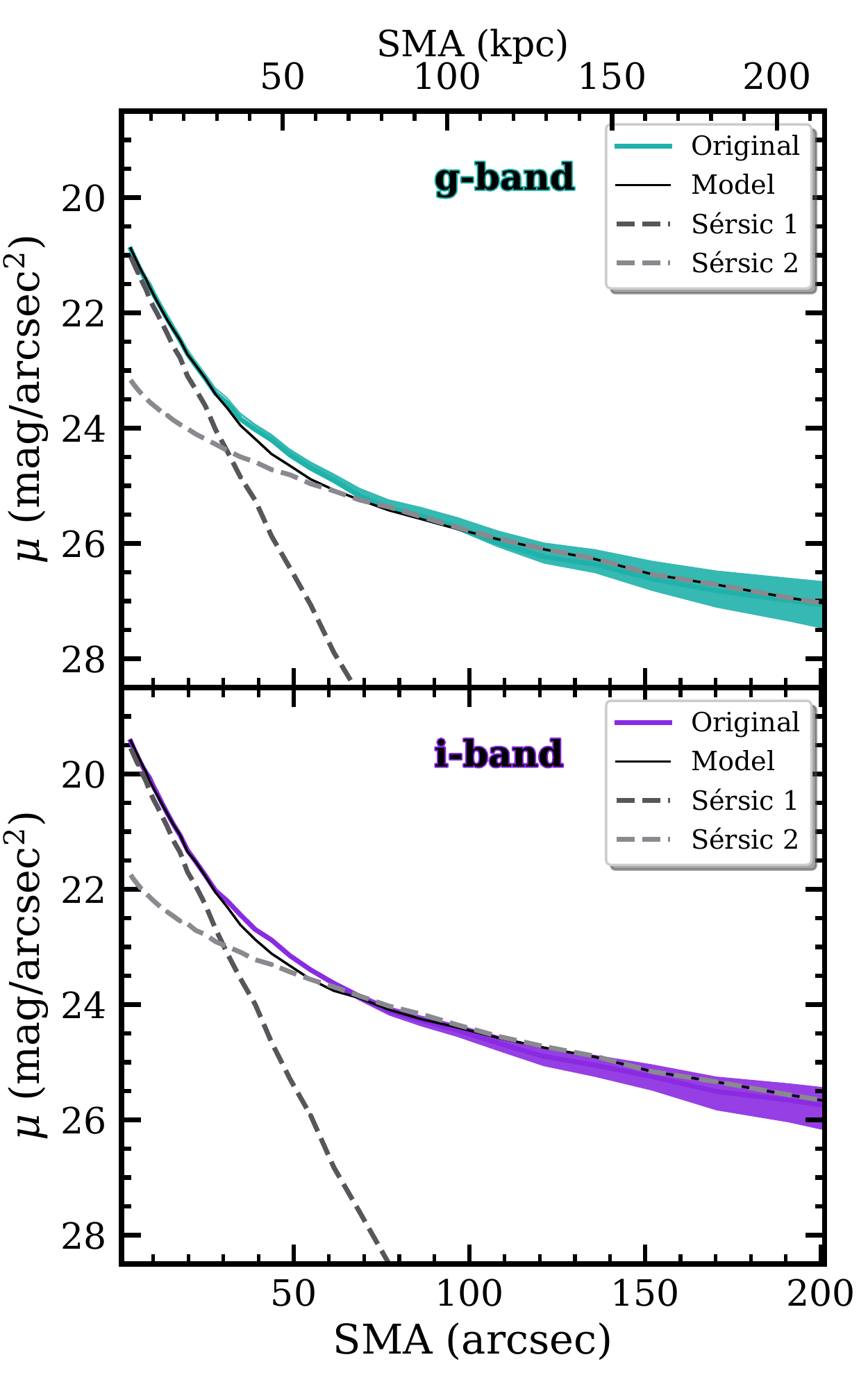}
\caption{\texttt{ellipse} 1-D profiles of the \g{} (top) and \ib{} (bottom) bands of the BCG + ICL of A85. The grey dashed lines correspond to the profiles of the two S\'ersic components fitted with GALFIT. The solid black line is the sum of both components. A double S\'ersic fit reproduces the light profile of A85 in both bands and the outer component, the ICL, dominates at around $\sim$ 60-70$\arcsec$.\label{fig:fit2d}}
\end{center}

\end{figure}

The 1-D surface brightness profiles obtained with \texttt{ellipse} for the double S\'ersic fits are shown in Fig. \ref{fig:fit2d}. As in Fig. \ref{fig:profiles}, the observed surface brightness profiles of the \g{} and \ib{} bands are shown in mint green and purple, respectively. The two different S\'ersic models (inner and outer) are shown with the dashed grey lines while the sum of both models is the solid black line. As in Fig. \ref{fig:fit_b4}, it can be seen that the outer component, the ICL, dominates at around $\sim$ 60-70$\arcsec$. 

The ICL fraction obtained using the outer S\'ersic model is given in Table \ref{tab:fractions}. We have also derived the fraction of BCG+ICL with respect to the total and the ratio between ICL and BCG+ICL.

\vspace{1cm}

The ICL fractions derived from assuming a double S\'ersic to describe BCG+ICL are higher than those from surface brightness cuts. This is expected because we are extrapolating the contribution of the diffuse light in the line of sight of the BCG which results in adding more ICL that surface brightness cuts cannot account for. Fig. \ref{fig:fit2d} shows that while the extended S\'ersic component begins to dominate at r $= 70$ kpc, a surface brightness limit of $\mu_g>26$ mag/arcsec$^2$ will measure all the light beyond $110$ kpc as ICL. 
Note that a surface brightness cut also accounts for diffuse light that is associated with the other galaxies in the cluster and might give a more complete picture of the formation of ICL in clusters. 

The fractions calculated in this Section include all the member galaxies of the cluster in our images. That is r = 0\fdg42 = 0.67$\times R_{200}$ \citep[where R$_{200} = $0\fdg63 = 2.42 Mpc,][]{Owers2017}. Other studies measure the ICL fraction in smaller radius, tipically $R_{500}$ \citep[e.g.,][]{Gonzalez2007}. This means that, in comparison, we are including more galaxies and therefore deriving a higher total luminosity for the cluster\footnote{While not adding almost any ICL.}. That yields to lower ICL fractions than in other studies with more limited field of views \citep[e.g.,][]{Burke2015, DeMaio2020}. For this reason, we have also calculated the fractions within $R_{500} = 1.2$ Mpc = $18\farcm7$ \citep{Ichinohe2015}. The fractions within $R_{500}$ can be found in Table \ref{tab:fractions2} in Appendix \ref{app:fractions}.

\section{Discussion}
In this work, we have used archival HSC data to explore the radial surface brightness and color profile of the BCG of A85 to a radius of $200\arcsec$ ($213$ kpc). We found that both the surface brightness and color profile become shallower beyond $70\arcsec$ ($75$ kpc), indicating that an extra component, the ICL, starts to dominate. In the following, we will discuss the implications of our results.

\subsection{The fraction of light in the ICL}

The ICL is a product of the interactions between galaxies within the cluster \citep[][]{Rudick2009}, therefore its fraction can provide information of the efficiency of those interactions, while the evolution of this component with time gives an estimation of their timescales. However, measuring the ICL fraction is difficult as the transition between BCG and ICL happens smoothly, making it hard to separate both components. In addition, studies use different bands and definitions for the ICL complicating direct comparison.

In general, the ICL fractions derived here using surface brightness cuts are in agreement with those in the literature for clusters at similar redshifts (although in different, adjacent, bands and surface brightness limits, e.g., \citealt{Krick2007}). Our ICL fraction at $\mu_g>26 $ mag/arcsec$^2$ is $\sim9.8\pm0.5\%$ (Table \ref{tab:fractions2}). This is in agreement with the median ICL fraction (using the same band and surface brightness cut) in \citet{Kluge2020b}: $13\pm13\%$. It is also in agreement with the $\sim11\%$ at $\mu_V>26.5$ mag/arcsec$^2$ derived in the simulations of \citet{Rudick2011}. 

Simulations show that $70\%$ of the stellar mass of the BCG is accreted \citep{Qu2017, Pillepich2018}. This means that most of the BCG is formed in a similar way to the ICL, and therefore they should be studied in order to understand the growth of BCGs. For this reason, we also measured the fraction of BCG+ICL over the total luminosity of the cluster, $f_{BCG+ICL}$. This fraction is $\sim46\%$ at $r<R_{500}$ in agreement with \citet{Gonzalez2007, Gonzalez2013} for clusters at similar redshifts.

The fraction of ICL over the BCG+ICL component, f$_{\mathrm{ICL}/\mathrm{BCG+ICL}}$ ($\sim 64\%$) indicates that most of the total light in the BCG+ICL system in A85 is in the ICL\footnote{More specifically, it is the stellar halo or envelope (bound to the BCG) + ICL (bound to the cluster) as we cannot distinguish between both components using imaging alone.}. This result agrees with the fractions from previous observations and simulations \citep[e.g.,][]{Gonzalez2005, Zibetti2005, Seigar2007, Canas2020, Kluge2020b}. In the simulations of \citet{Conroy2007}, similar fractions are achieved if all the stars from disrupted satellites end up in the ICL (their Fig. 4). These results from simulations, coupled with the observed mild evolution in mass of BCGs \citep[e.g.,][]{Whiley2008, Collins2009, Lidman2012, Oliva2014, Bellstedt2016}, suggests indicate that a significant fraction of the mass of infalling satellites goes to the stellar halo + ICL instead of adding a significant fraction of mass to the BCG \citep[e.g.,][]{Laporte2013, Contini2018}.

\subsection{Stellar populations of the BCG}\label{sec:disc_sp}

Studying the colors of the ICL in clusters allows us to infer the properties of the progenitor galaxies from which the ICL accreted its stars and, consequently, the mechanisms at play in the formation of this component.

In Section \ref{sec:profiles}, we presented the surface brightness radial profiles of the BCG + ICL for the \g{} and \ib{} bands. Both surface brightness profiles show a flat region in the inner $10\arcsec$ ($11$ kpc), denoting the presence of a core \citep[e.g.,][]{Lopez-Cruz2014, Madrid2016}. In the same way, the measured color profile is flat in the inner $10\arcsec$, indicating that the stellar populations in this region are well mixed. \citet{Mehrgan2019} used MUSE data to infer the stellar kinematics of this BCG finding that this central region hosts a supermassive black hole with a mass of $4.0\pm0.8 \times10^{10} M_{\odot}$. They concluded that the BCG of A85 is a result of the merger of two cored early-type galaxies. 

Beyond $10\arcsec$, the surface brightness profiles follow a \citet{Sersic1968} profile down to $\sim70\arcsec$ ($\sim 75$ kpc). At the same time, the color profile shows a negative gradient from $g-i = 1.45$ to $g - i \approx 1.25$. The central flattening and subsequent gradient in color is also observed in the integral field spectroscopy observations of A85 in \citet{Edwards2020}. They find that out to $30$ kpc the metallicity of the galaxy shows the same behaviour as our color profile: a flattening in the inner $\sim10$ kpc ($\sim10\arcsec$), followed by a decrease to $\sim30$ kpc ($28\arcsec$).

At $\sim70\arcsec$, the surface brightness profiles depart from the S\'ersic fit (top panel in Fig. \ref{fig:fit_b4}). This corresponds to where the isophotes show a boxy shape (indicated by the gray vertical line in Fig. \ref{fig:fit_b4}). Simulations suggest that boxyness is the result of a past dry merger event \citep[e.g.,][]{Naab2006}. In addition, at this radius, the color profile becomes shallower. These pieces of evidence point to an extra component originating from accreted stars: the ICL\footnote{ICL + stellar halo.}.

It is not possible to disentangle between stellar age and metallicity using only one color. Previous deep observations of clusters of galaxies show clear radial gradients in colors \citep[e.g.,][]{Williams2007, MT14, MT18, DeMaio2015, DeMaio2018, Mihos2017, Iodice2017} indicating radial gradients in metallicity while the ages of the ICL in nearby systems are old \citep[$>10$ Gyr, e.g.,][]{Williams2007, Coccato2010}. This is consistent with \citet{Edwards2020}, who only found a very mild decrease in age to $30$ kpc for A85, from $\sim15$ to $10$ Gyr. Therefore at $<30$ kpc, the color profiles likely mostly trace changes in metallicity. However, we cannot test here whether the decrease in age becomes significant beyond $30$ kpc. 

The shape of the color profile is reminiscent of the three different regions found in the metallicity profile of M87 in \citet{Montes2014a} (see also \citealt{Coccato2010}). In M87, the metallicity gradient becomes shallower in the outer parts of the galaxy. This is the consequence of the mixing of the stellar populations of the accreted galaxies. This also appears to be the case for A85 and is supported by the change in the slope of the surface brightness profiles, where the outer parts of the galaxy (the ICL) are built via the accretion of satellite galaxies.  

\begin{figure*}
\begin{center}
\includegraphics[scale = 1.2]{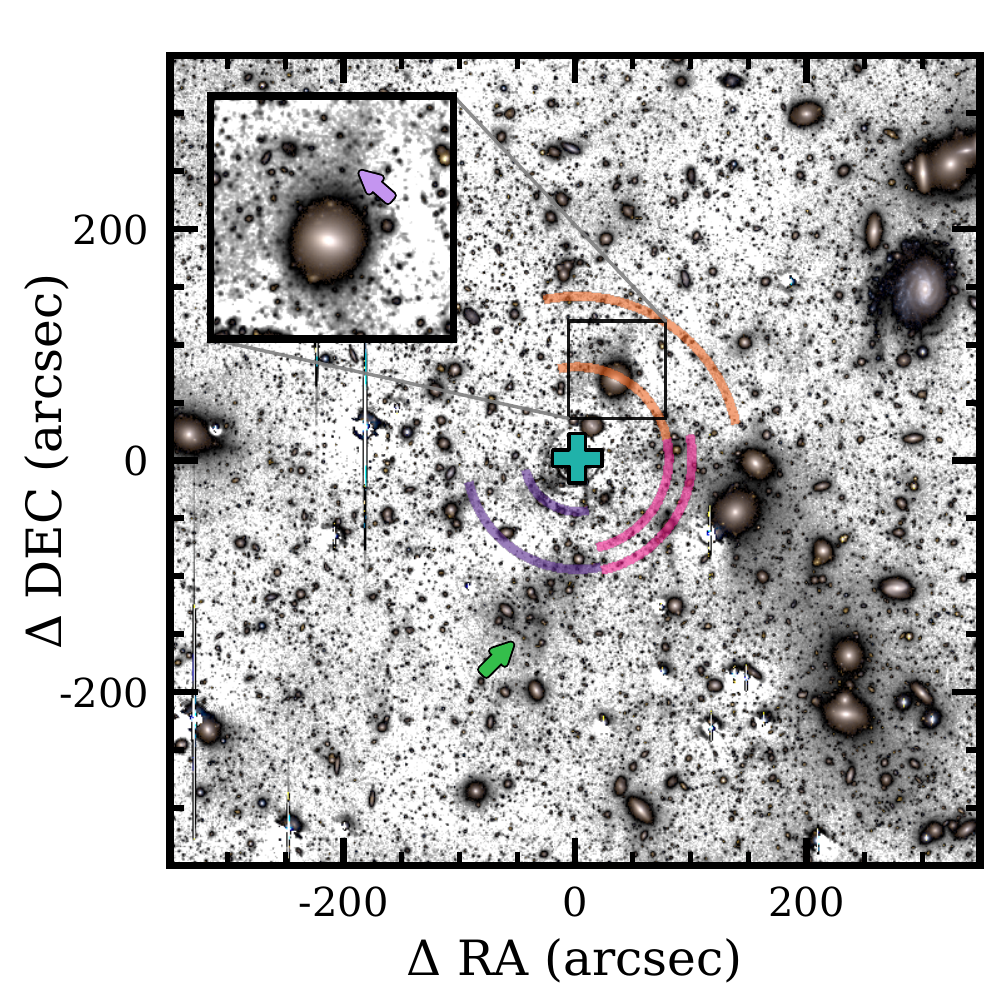}
\caption{Inner $700\arcsec\times700\arcsec$ RGB image of A85. The \texttt{ellipse} model obtained in Sec. \ref{sec:profiles} has been subtracted. The teal cross marks the centre of the BCG while the green arrow marks a collection of galaxies and diffuse light to the South of the BCG. The colored arcs mark the area where there is flattening in the color profiles derived in different directions as seen in Fig. \ref{fig:quarters}. The inset shows a zoom-in into a galaxy that presents a faint tail towards the North as indicated by the purple arrow. The North (orange) and West (magenta) arcs are highlighting diffuse light associated with galaxies interacting with the BCG.} \label{fig:res}
\end{center}

\end{figure*}

In Section \ref{sec:color}, we derived color profiles of the BCG+ICL in four different 90 deg-wide sections, finding that the Southern color profile between $50\arcsec$ to $100\arcsec$ ($53$ to $106$ kpc) is redder than the other profiles ($\sim1.3$, Fig. \ref{fig:quarters}). Similarly, the North and West profiles become flat between $80\arcsec$ to $130\arcsec$. 

To explore whether there is any evidence of infalling material that might be causing the flattening of the profiles, we subtracted the \texttt{ellipse} models from the image for both bands to enhance any signs of interactions or asymmetries. In Fig. \ref{fig:res}, we show the inner $700\arcsec\times700\arcsec$ region of A85 with the  model generated from the \texttt{ellipse} fits subtracted. 
We have drawn arcs to demarcate the areas in the image corresponding to the flattening of the color profiles, color-coded according to the direction of the corresponding profile in Fig. \ref{fig:quarters}. 
Towards the North, we found a faint tail associated with a large satellite galaxy (zoomed-in in Fig. \ref{fig:res} and marked with a purple arrow). The presence of this faint tail might explain the sudden reddening of the Northern color profile at $\sim130\arcsec$ in Fig. \ref{fig:quarters}. As there are no other signs of disturbance, this galaxy is probably just starting to interact with the BCG. To the West, there is some diffuse light associated with two galaxies that are likely interacting with the BCG. Even with our careful and conservative masking, the diffuse light associated with these interactions might be contaminating our profiles.

In addition, there is a collection of galaxies between $\sim75\arcsec$ and $\sim160\arcsec$ ($80$ to $170$ kpc) South of the BCG with associated diffuse light. This structure is marked with a green arrow in Fig. \ref{fig:res}. These galaxies appear to be interacting with each other rather than with the BCG. The color of the diffuse light of the region, with the galaxies masked, is $g-i = 1.24\pm0.15$. However, this structure is not associated to any signature in the Southern color profile (lies outside the purple arcs in Fig. \ref{fig:res}).

To investigate whether these diffuse structures could be biasing our results, we repeated the \texttt{ellipse} fitting and color profile derivation but, in this case, generously masking all of these structures (North, South and West). The resulting color profile and the $4th$ Fourier harmonic, B4, are shown in Fig. \ref{fig:color_test} in Appendix \ref{app:color_test}. We also plotted the original color profile in dark blue for reference. The new color profile is compatible with the original. The gradient between $70\arcsec$ to $200\arcsec$ ($75$ to $213$ kpc) is now $-0.07\pm 0.04$, slightly steeper but compatible within errors with the gradient derived in Sec. \ref{sec:color} ($-0.06\pm0.04$). The boxyness is still present. Therefore, the diffuse light associated with these interactions is not producing the change of gradient nor the boxyness observed. 


The lack of any obvious tidal feature related to the color flattening towards the South means that any tidal feature has had time to smear, only emerging here in the color profile. Given its preferential position to the South, it did not have enough time to mix with the rest of the galaxy. The orbital period for a star around this BCG in the radial range from $50\arcsec$ to $100\arcsec$ ($53$ to $106$ kpc, the approximate range of the flattening in the Southern color profile) is between $1.5$ to $3.7$ Gyr. The calculations of the orbital period are described in Appendix \ref{app:orbital}. In the simulations of \citet{Rudick2009}, streams found in the core of the cluster are quickly destroyed by the strong, evolving tidal field of the cluster with timescales of $\lesssim1$ Gyr. Therefore, for a star not to have orbited the galaxy but any stream to be smeared, we suggest that this interaction likely happened a few Gyrs ago.

\subsubsection{The ellipticity profile and the ICL as a tracer of dark matter}

A significant anisotropy in the orientation of the orbits of the progenitors of the ICL will produce an elongation in the ICL distribution. This elongation, i.e. ellipticity, is expected to increase with increasing radius up to the value of the original distribution, i.e. the cluster distribution. 
The ellipticity of the diffuse light of A85 increases with radius up to a value of $\sim 0.55$ at $\sim200\arcsec$ ($\sim 213$ kpc), for \g{} and \ib{} (Fig. \ref{fig:profiles}). However, the PA does not change significantly with radius, i.e., inner and outer components are aligned.

This increase in ellipticity with radius was also observed in this cluster by \citet{Kluge2020}. The same trend with radius has been measured in other massive galaxies and clusters \citep[e.g.,][]{Gonzalez2005, Tal2011, Huang2018, Kluge2020, Kluge2020b}. The ellipticities of the diffuse light in these systems tend to the typical values for the distribution of galaxies within clusters \citep{Shin2018}. When fitting a double S\'ersic model to the 2-D distribution of light of the BCG+ICL, we also find that the ellipticity of the outer component, the ICL, has a higher ellipticity ($\sim 0.5$) than the inner component, the BCG ($\sim0.2$).

The value of the ellipticity at large radii derived here is consistent with the axis ratio measured for A85 using weak-lensing modeling by \citet{Cypriano2004}. That is, the ICL has the ellipticity of the dark matter halo of the cluster. These results agree with the picture proposed in \citet{MT19} that the ICL is a good luminous tracer of the dark matter distribution in clusters of galaxies.

\subsection{The buildup of the ICL of A85}

The change in slope of the surface brightness profile of the BCG, the boxyness of the isophotes and the change in the slope of the color gradient at a radius of $\sim70\arcsec$ ($\sim 75$ kpc) suggests strongly that the BCG and ICL can be considered as distinct stellar components with different assembly histories, and that the accreted component (ICL) starts to dominate at that radius. Integrated light spectroscopy \citep[e.g., ][]{Dressler1979} and planetary nebulae kinematics \citep[e.g., ][]{Arnaboldi1996} of nearby clusters, show that the radial velocity dispersion increases with radius to reach the value of the velocity dispersion of the galaxies in the cluster \citep{Longobardi2018}. That means that the stars forming the ICL are following the potential of the cluster rather than the potential of the BCG.
We can conclude that the radius where the potential of the A85 cluster begins to dominate is $\sim70\arcsec$ ($\sim75$ kpc, see Fig. \ref{fig:fit2d}). Previous works have also shown that, in massive clusters ($10^{14-15}M_{\odot}$), BCGs tend to show this break radius at around $60-80$ kpc \citep[e.g., ][]{Zibetti2005, Gonzalez2005, Seigar2007, Iodice2016}.

We can calculate an approximate mass of the progenitor of the merger using the color of the Southern profile. If the average color of the reddening in the Southern profile is around $g-i = 1.3$ (Fig. \ref{fig:quarters}), and assuming an age of $10$ Gyr, the metallicity of the progenitor would be [Z/H] $= -0.013$ \citep[using the models of][]{Vazdekis2016}, i.e. slightly subsolar metallicity. Using the mass-metallicity relation from \citet{Gallazzi2005}, this corresponds to a galaxy of $\sim3\times 10^{10} M_\odot$. The galaxies towards the North and West that are interacting with the BCG have masses of the order of $\sim7\times10^{10} M_\odot$ \citep{Owers2017}. This is in agreement with observations in other clusters \citep{MT14, MT18, Morishita2017, DeMaio2018} and with simulations \citep{Purcell2007, Cui2014, Contini2014, Contini2019}. These studies conclude that galaxies of $\sim10^{10} M_\odot$ are the main contributors to the ICL in massive clusters of galaxies.

\section{Conclusions}
In this work, we have presented deep observations in the \g{} and \ib{} bands of the central $52\arcmin \times 52\arcmin$ of the cluster Abell 85, obtained with Hyper Suprime-Cam on the Subaru Telescope. The surface brightness limits reached are $30.1$ and $29.7$ mag/arcsec$^2$ ($3\sigma$, $10\times10$ arcsec$^2$), for the \g{} and \ib{} bands, respectively.
Taking advantage of the depth of these images, we are able to study the diffuse light of this cluster down to $200\arcsec$ ($213$ kpc) from the BCG. At $\sim70\arcsec$ ($\sim75$ kpc), the surface brightness profiles become shallower and the isophotes show a boxy shape, strongly indicating the presence of an accreted component: the ICL. In addition, at the same radius the color profile becomes shallower, a consequence of the mixing of the stellar populations of the accreted galaxies.

Furthermore, the color profile towards the North, West and South of the BCG show a redder color compared to the other profiles as if there is remaining material in that direction from a merger that happened a few Gyrs ago. This work shows that even short exposure times ($\sim30$ mins) on large telescopes can unveil the assembly history of clusters of galaxies.

The results presented in this work show the extraordinary power of ground-based observatories to address the origin and evolution of the ICL. In the future, the LSST will be able to provide deep multi-wavelength observations of the southern sky allowing the study of the ICL in a range of cluster masses and redshifts \citep{Montes2019, Brough2020}. However, as demonstrated in this work, careful data processing techniques are crucial in order to take the maximum benefit from the upcoming data.

\acknowledgments
We thank the referee for constructive comments that helped to improve the original manuscript. MM would like to thank Ra\'ul Infante-S\'ainz and Nacho Trujillo for their very useful comments on the data reduction and Alexandre Vazdekis for his comments on stellar populations. SB acknowledges funding support from the Australian Research Council through a Future Fellowship (FT140101166). Based on data collected at Subaru Telescope and obtained from the SMOKA, which is operated by the Astronomy Data Center, National Astronomical Observatory of Japan. This research includes computations using the computational cluster Katana supported by Research Technology Services at UNSW Sydney.

\facilities{Subaru Telescope}

\software{Astropy \citep{Astropy2018},  
          \sextractor{} v2.19.5 \citep{Bertin1996},
          \swarp{} v2.38.0 \citep{Bertin2002}, 
          \scamp{} v2.0.4 \citep{Bertin2006},
          \texttt{photutils} v0.7.2 \citep{Bradley2019},
          \texttt{pillow} \citep{pillow2020},
          \texttt{ellipse} \citep{Jedrzejewski1987},
          GALFIT \citep{Peng2002}}

\newpage
\appendix

\section{Effect of instrument rotation on the flat-fields}\label{app:rot_flat}

In Sec. \ref{sec:flat}, we discussed that we used HSC-SSP images of adjacent nights to the observations of A85 in order to derive a sky flat. However, in the case of the \ib{} band, using the images of adjacent nights resulted in significant background substructure in the individual CCDs, and consequently, a global structure in the individual frames. The source of this structure seems to be related to the rotation angle of the instrument (`INR-STR' in the header) being considerably different from that of the observed images ($<$INR-STR$>$ = $-5$ in the HSC-SSP Wide images compared to $<$INR-STR$>$ = $124$ for A85).

\begin{figure*}
\begin{center}
\includegraphics[width=0.8\textwidth]{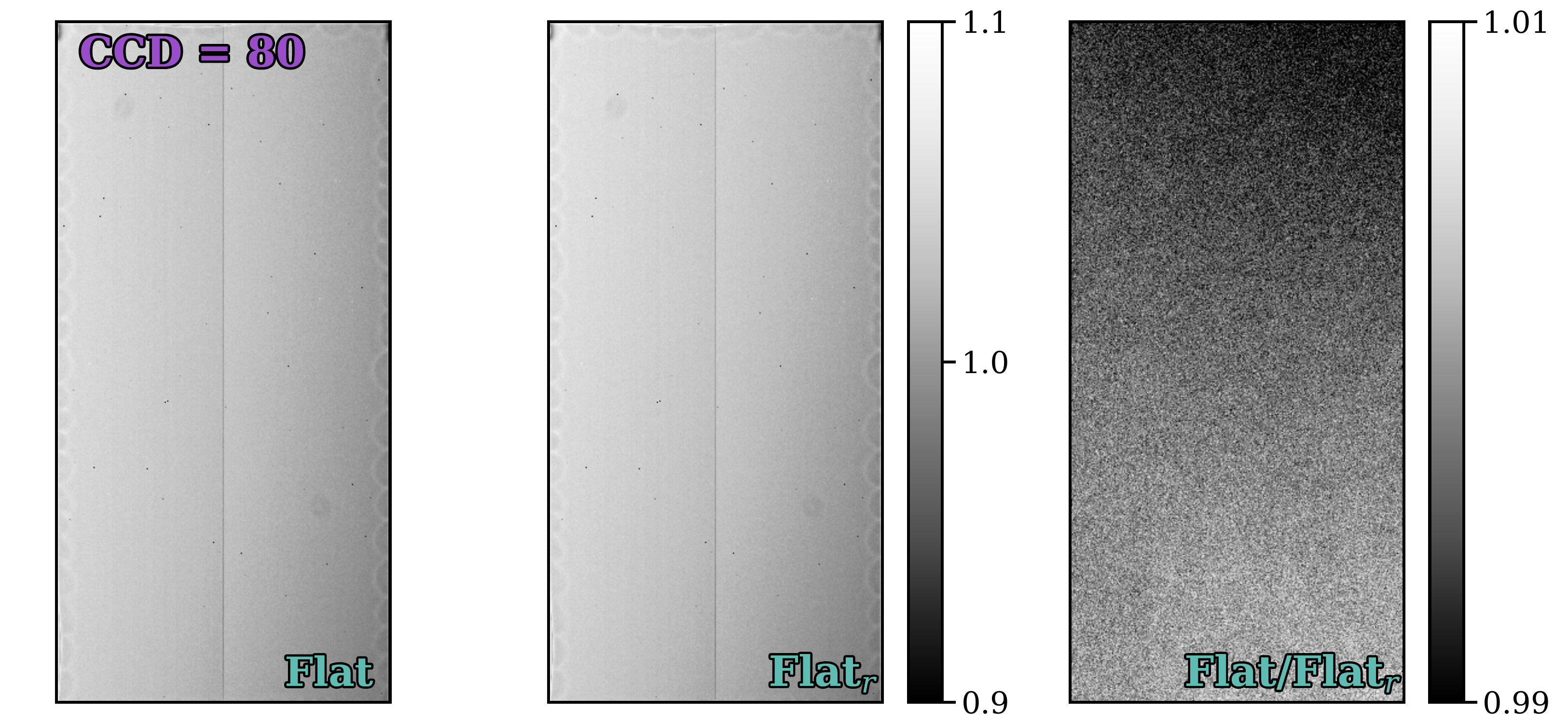}
\caption{Effect of rotation angles of the HSC (`INR-STR') on the flat-fields derived for CCD = 80 in the \ib{} band. Left panel: Flat derived using the images taken from adjacent nights to the images of A85 ($<$INR-STR$>$ = $-5$). Middle panel: Flat derived using the images with similar rotation angles as the A85 science images ($<$INR-STR$>$ = $114$). Right panel: Ratio of the two flats, Flat/Flat$_r$, where a gradient of $\sim1\%$ across the CCD can be seen. For reference $<$INR-STR$>$ of the A85 images is $124$.\label{fig:flat_rot}}
\end{center}
\end{figure*}

To test this hypothesis, we downloaded images from the SMOKA archive where `INR-STR' was close to the angle of the A85 \ib{} band images. As it was difficult to find the same rotation angles as the A85 images, we downloaded images with angles between 100 and 140. The average rotation angle of these images are $<$INR-STR$>$ = 114. The dates when those images were taken are listed in Sec. \ref{sec:flat}. 
In Fig. \ref{fig:flat_rot}, we show a comparison of the two different flats derived for CCD number 80\footnote{Map of the CCD arrangement here: \url{https://hsc.mtk.nao.ac.jp/pipedoc/pipedoc_4_e/_images/CCDPosition_20140811-1.png}}. The flat derived from the images from adjacent nights is shown in the left panel, labelled as \emph{Flat}. The middle panel shows the flat derived from the images with a median instrument rotation close to the A85 images, labelled as \emph{Flat$_r$}. The right panel of Fig. \ref{fig:flat_rot} is the ratio of the two flats; \emph{Flat} divided by \emph{Flat$_r$}. The presence of a significant gradient across the CCD can be seen. This gradient is of the order of $\sim1\%$. 

\begin{figure*}
\begin{center}
\includegraphics[width=0.9\textwidth]{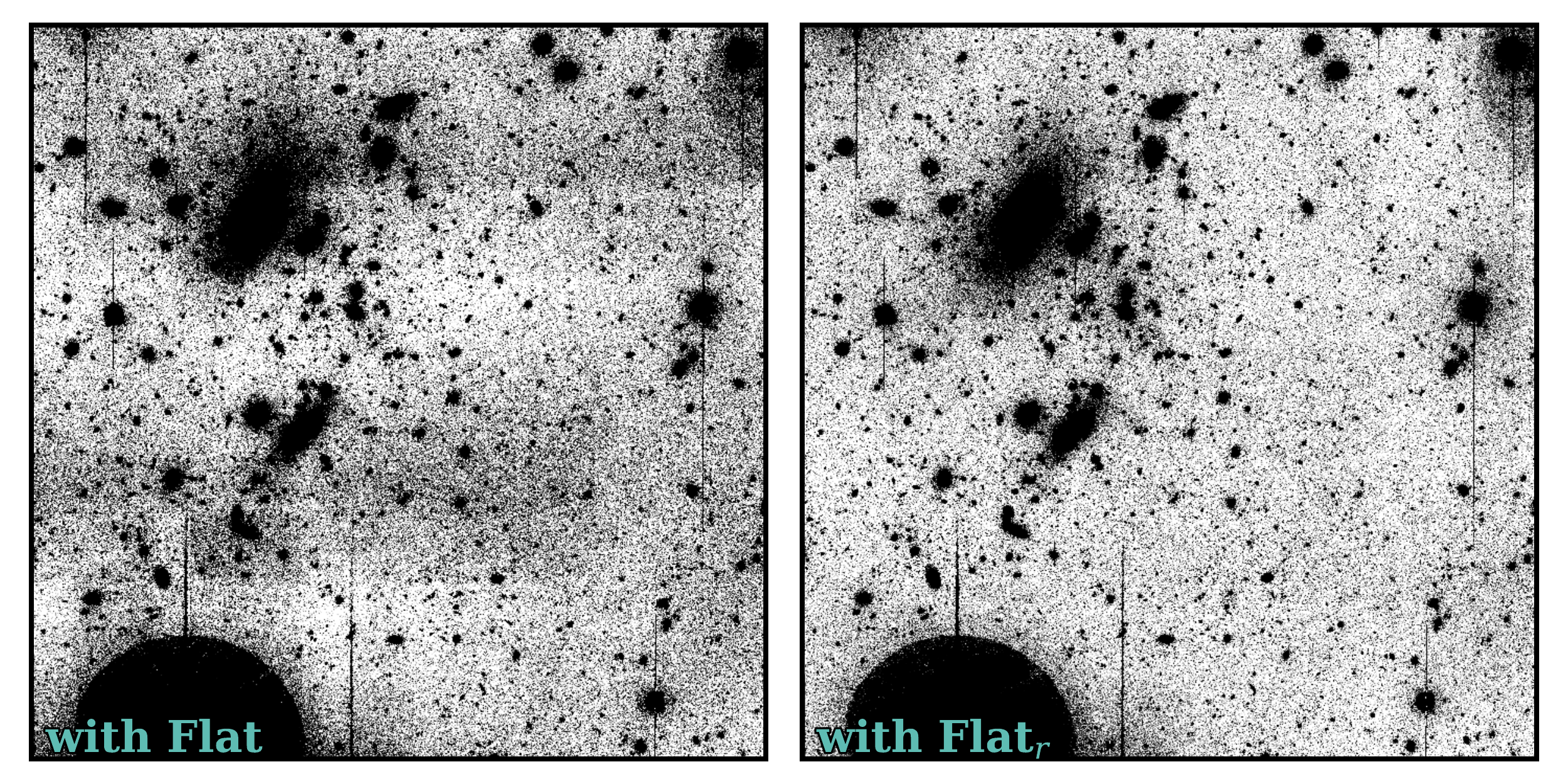}
\caption{Final \ib{} band co-added image using the flat-field frames from science images of adjacent nights to the A85 observations (`with Flat', left panel) and using the flat-field frames from science images with the same rotation as A85 (`with Flat$_r$', right panel). The left image shows inhomogeneities caused by the inaccuracy of the flat-field correction.}\label{fig:flat_im}
\end{center}

\end{figure*}

In Fig. \ref{fig:flat_im}, we show the comparison of the final co-added images for the \ib{} band using the flats derived using science images of adjacent nights (labelled: with Flat, left panel) and using the flats obtained with the science images with the same rotation as the A85 image (the final image used in this work labelled: with Flat$_r$, right panel). In the left image we can see inhomogeneities caused by the poor flat-field correction to the individual CCDs. 

\section{Example of the custom-made flat-field}\label{app:flat_ex}
\begin{figure*}
\begin{center}
\includegraphics[width=0.9\textwidth]{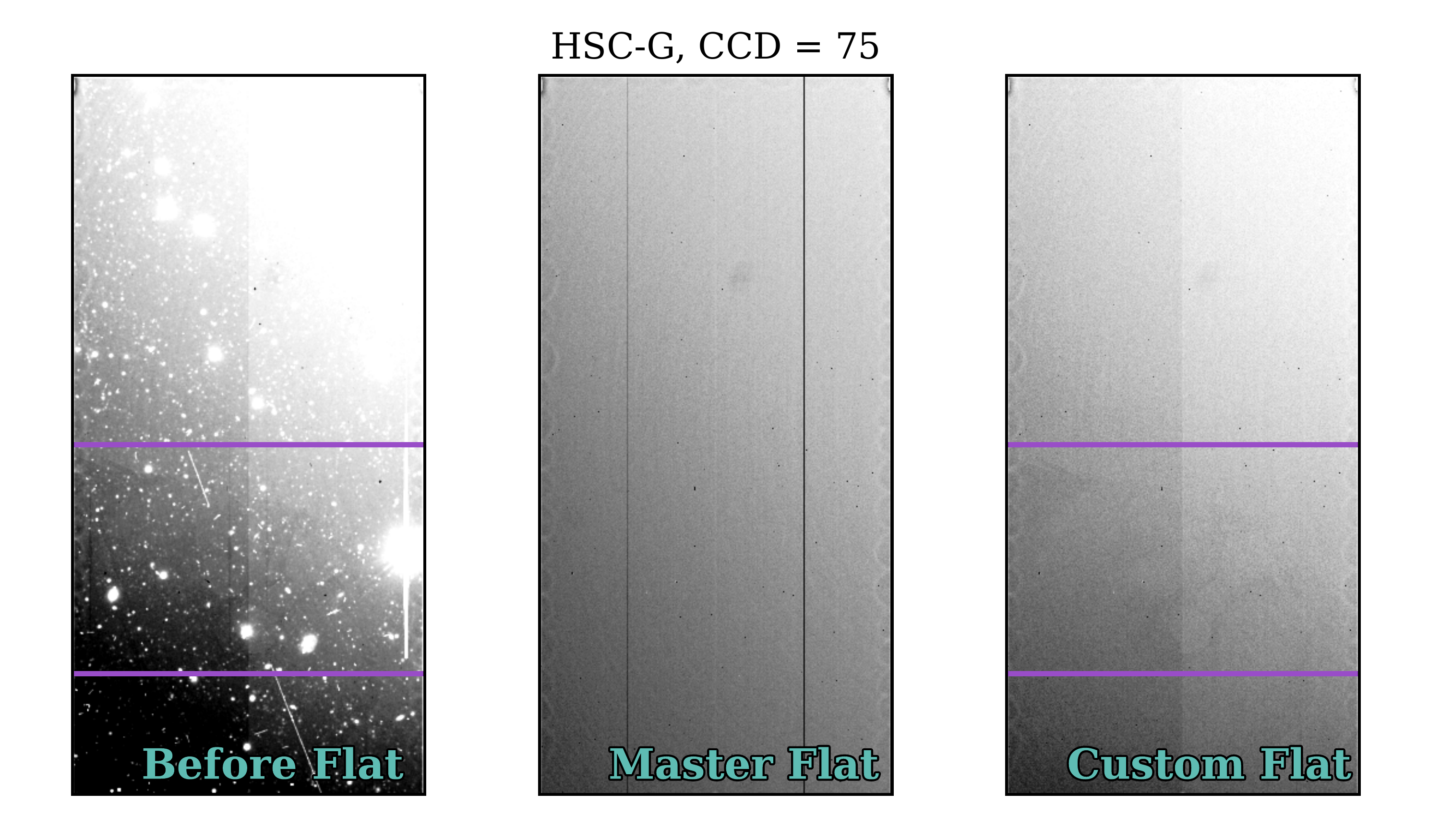}
\caption{Left panel: Image from CCD 75 processed before flat-field correction. Middle panel: Master flat-field for CCD 75 from 2014-09-17. Right panel: Custom flat derived in this work. The purple lines highlight the structure present in both the image and custom flat, but not in the HSC master flat.}\label{fig:flate_subs}
\end{center}

\end{figure*}

An accurate flat-field correction is crucial to minimise errors in low surface brightness science, especially in extended and diffuse objects such as the ICL. For this reason, we derived the flats from science observations instead of using the HSC master flats as inhomogeneities in the illumination can introduce gradients in our images. During this work, we found that the HSC master flat from CCD 75 does not contain a feature that is present in the data (indicated by purple lines in Fig. \ref{fig:flate_subs}). We do not know the reason for this discrepancy. However, the master flat does seem to reproduce all other features seen in the CCD image. 

\section{S\'ersic fits 1D and 2D}\label{app:sersics}
\begin{figure}
\begin{center}
\includegraphics[scale = 0.6]{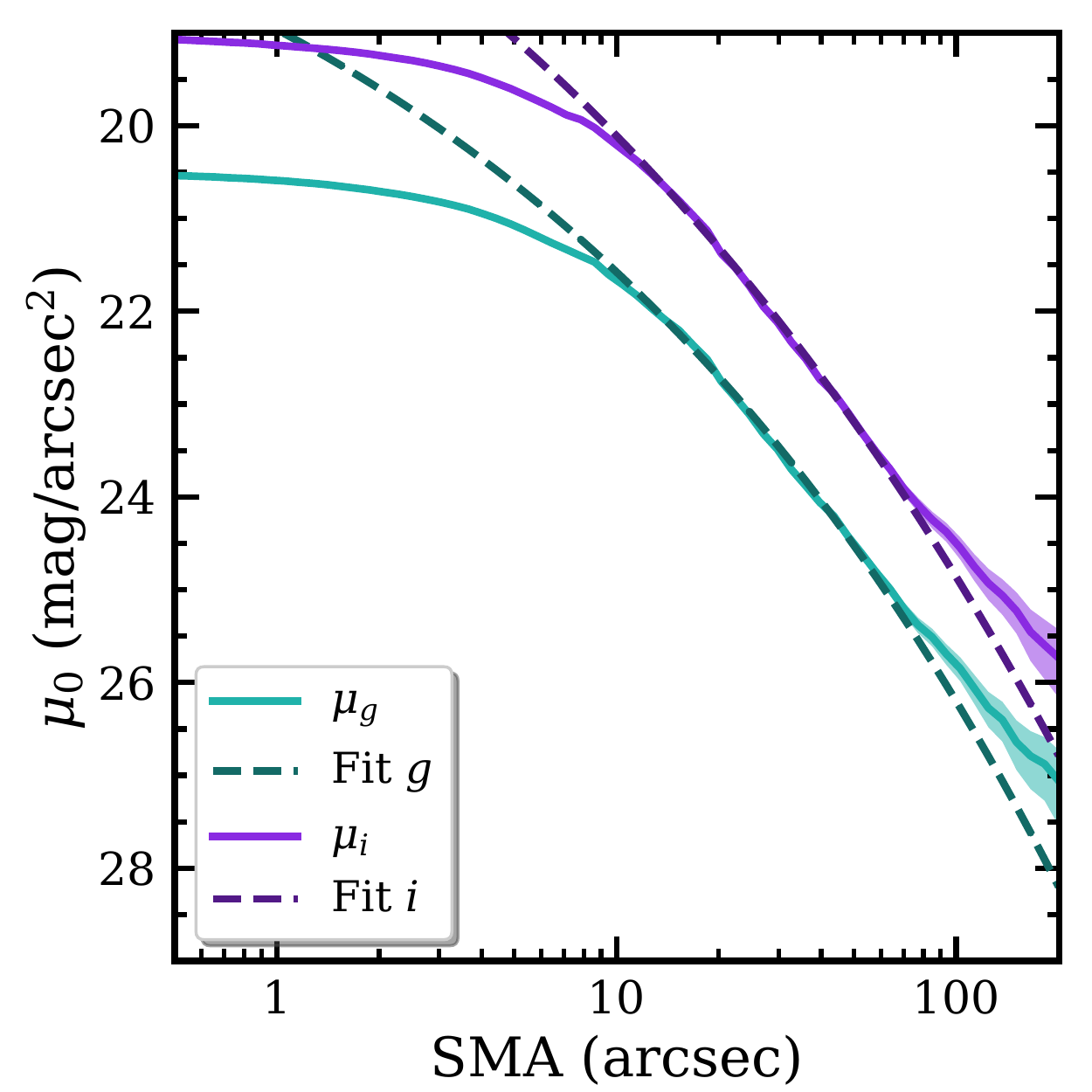}
\caption{1-D single S\'ersic fits to the BCG+ICL profile of A85.}\label{fig:fit_sersic}
\end{center}

\end{figure}

In Fig. \ref{fig:fit_sersic}, we show the single S\'ersic fits to the radial surface brightness profiles of A85, for the \g{} (green) and \ib{} (purple) bands. The best fit effective radius ($r_{eff}$) and S\'ersic index (n) parameters are listed in Table \ref{tab:sersic}, for both bands.

\begin{deluxetable}{ccc}[h]
\tabcolsep=0.4cm
\tablecaption{\label{tab:sersic}
	Parameters of the single S\'ersic fits for the surface brightness profiles of A85}
\tablehead{Band	       	 & $r_{\mathrm{eff}}$ [arcsec]	  & n }
\startdata
g         & $39\pm6$ & $4.0\pm0.7$ \\
i         & $36\pm4$ & $4.9\pm0.8$ 
\enddata
\end{deluxetable}

\begin{deluxetable}{ccccccccccccc}[h]
\tabcolsep=0.17cm
\tablecaption{\label{tab:sersic2d}
	 Parameters from the double S\'ersic fit from GALFIT}
\tablehead{Band	  & m$_1$  & $r_{\mathrm{eff},1}$  & n$_1$  & $\varepsilon_1$ & PA$_1$ & m$_2$& $r_{\mathrm{eff},2}$  & n$_2$  & $\varepsilon_2$ & PA$_2$   \\
 & [mag] &[arcsec] & & & & [mag]& [arcsec] &  & &}
\startdata
g    &  $14.37\pm 0.01$  & $15.02\pm0.01$ & $1.08\pm0.01$ & $0.20\pm0.01$ & $54.40 \pm0.02$ & $13.49\pm0.01$  & $173.0\pm0.2$ & $2.14\pm0.01$ & $0.49\pm0.01$ & $58.63\pm0.02$ \\
i    & $12.96 \pm 0.01$    & $14.29\pm0.01$ & $1.04\pm0.01$ & $0.19\pm0.01$  & $54.16\pm0.02$ & $12.20\pm0.01$  & $172.6\pm0.2$ & $2.18\pm0.01$ & $0.53\pm0.01$ & $58.82\pm0.02$
\enddata
\end{deluxetable}

\section{Color profiles of A85 after masking diffuse structures}\label{app:color_test}
\begin{figure}
\begin{center}
\includegraphics[scale = 0.6]{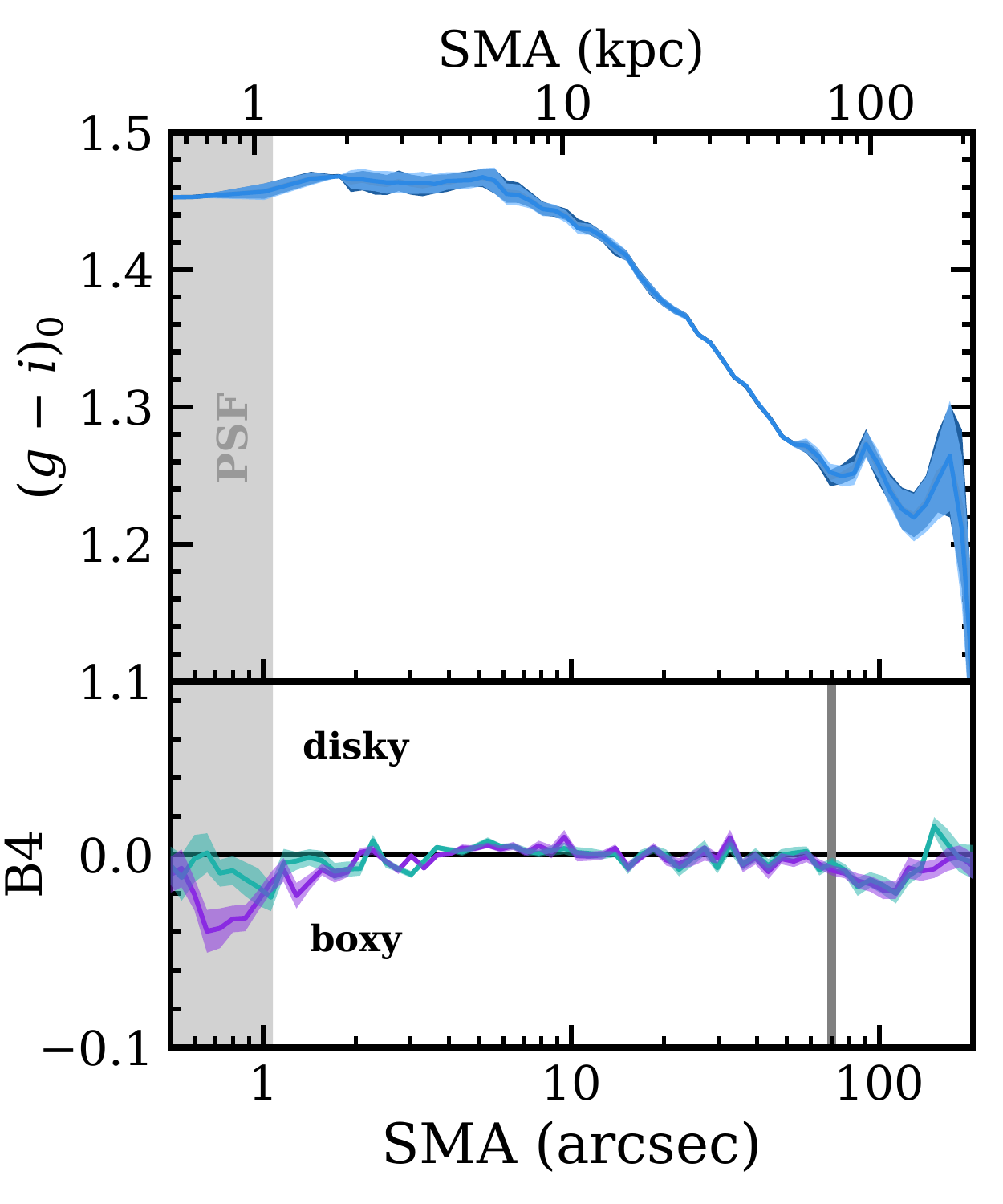}
\caption{Color profile and B4 coefficients as a function of SMA after masking the collection of galaxies to the South of the BCG. We do not find any significant difference within errors with respect to the results from Sec. \ref{sec:color} (in dark blue).}\label{fig:color_test}
\end{center}

\end{figure}

In Sec. \ref{sec:disc_sp}, we discussed the existence of diffuse structures North, South and West of the BCG with associated diffuse light that might be causing the flattening of the color profile. In Fig. \ref{fig:color_test}, we show the $g-i$ color profile (top panel) and the B4 coefficient (bottom panel) with these structures masked. We do not find any significant change within errors.

\section{Calculation of the orbital period around A85}\label{app:orbital}

To estimate the orbital period ($T$) of a star at a given radius ($R$) around A85 we used Kepler's third law:
\begin{equation}
    T = 2\pi\sqrt{\frac{R^{3}}{GM_R}}
\end{equation}
where $M_R$ is the mass of the BCG of A85 inside a radius $R$ and $G$ is the gravitational constant \citep{Binney1987}. 

In order to calculate $M_R$ for A85, we followed the prescription in \citet{Bell2003} to calculate the mass-to-light ratio ($M/L$) as a function of the $g-i$ color, assuming a \citealt{Salpeter1955} IMF. The radius is computed as the elliptical distance to the BCG, where the morphological parameters (ellipticity and PA) are the median values from the \texttt{ellipse} isophotes excluding the inner $10\arcsec$: 0.37 for the ellipticity and 56 deg for the PA. 
The expression we have used to estimate the $M/L$ in the \g{} band is:
\begin{equation}
    \log(M/L_g) = -0.379 + 0.914\times (g-i)
\end{equation}
from Table 7 in \citet{Bell2003}. The color used is the median $g-i$ color inside a radius $R$, $k$-corrected and corrected by the extinction of the Milky Way.

\section{ICL fraction at R$<$R$_{500}$} \label{app:fractions}
In Table \ref{tab:fractions2}, we list the ICL fractions inside $R_{500}$ (= 1.2 Mpc = 18\farcm7). We also list the BCG+ICL fraction with respect to the total luminosity of the cluster and the ICL fraction with respect to the BCG+ICL. 

\begin{deluxetable*}{lcccc}[t]
\tabcolsep=0.4cm
\tablecaption{\label{tab:fractions2}
	ICL fraction (\%) for A85 within r$<R_{500}$}
\tablehead{ \multicolumn{5}{c}{Surface brightness cuts } \\
& $26<\mu<29.5$ &  $26.5<\mu<29.5$ & $27<\mu<29.5$ & $27.5<\mu<29.5$ \\
\multicolumn{5}{c}{[mag/arcsec$^2$]}}
\startdata
$f_{\mathrm{ICL}}(g)$ & $9.8\pm0.5$ & $7.0\pm0.8$ & $4.6\pm1.0$ & $2.7\pm1.0$ \\
$f_{\mathrm{ICL}}(i)$ & $3.6\pm0.8$ & $2.2\pm0.8$ & $1.4\pm0.8$ & $0.7\pm0.8$ \\
\hline
\hline
\multicolumn{5}{c}{2D fit}\\
& \multicolumn{2}{c}{\g}  & \multicolumn{2}{c}{\ib}\\
\hline 
$f_{\mathrm{ICL}}$ &\multicolumn{2}{c}{$29.9\pm1.0$} & \multicolumn{2}{c}{$30.3\pm1.0$} \\
$f_{\mathrm{BCG+ICL}}$ &\multicolumn{2}{c}{$45.4\pm2.0$} &\multicolumn{2}{c}{$47.7\pm2.0$} \\
$f_{\mathrm{ICL}/\mathrm{BCG+ICL}}$ &\multicolumn{2}{c}{$66.1\pm2.2$}&\multicolumn{2}{c}{$63.7\pm2.2$}\\
\hline
\enddata
\end{deluxetable*}



\bibliography{icl_a85}{}
\bibliographystyle{aasjournal}



\end{document}